\documentstyle[epsfig,aps,prb]{revtex}
\def\beq{\begin{equation}}
\def\eeq{\end{equation}}
\def\beqa{\begin{eqnarray}}
\def\eeqa{\end{eqnarray}}
\tightenlines
\begin{document}
\title{Nonlinear Couette flow in a low density granular gas\footnote{Suggested 
running title: Couette flow in a granular gas}}
\author{M. Tij,\footnote{D\'{e}partement de Physique, Universit\'{e} Moulay 
Isma\"{\i}l,
Mekn\`{e}s, Morocco}
E. E. Tahiri,\footnote{D\'{e}partement de Math\'ematique, Universit\'{e} Moulay
Isma\"{\i}l,
Mekn\`{e}s, Morocco}
J. M. Montanero,\footnote{Departamento de Electr\'onica e Ingenier\'{\i}a 
Electromec\'anica,
Universidad  de  Extremadura, 
E-06071 Badajoz, Spain
}
V. Garz\'o,\footnote{
Departamento de F\'{\i}sica, Universidad  de  Extremadura, 
E-06071 Badajoz, Spain
}
A. Santos,$^{\dag\dag}$ and J. W. Dufty\footnote{Department of Physics, 
University of 
Florida, Gainesville, FL
32611}}
\date{\today}

\maketitle

\begin{abstract}
A model kinetic equation is solved exactly for a special stationary state
describing nonlinear Couette flow in a low density system of inelastic
spheres. The hydrodynamic fields, heat and momentum fluxes, and the phase
space distribution function are determined explicitly. The results apply for
conditions such that viscous heating dominates collisional cooling,
including large gradients far from the reference homogeneous cooling state.
Explicit expressions for the generalized transport coefficients (e.g.,
viscosity and thermal conductivity) are obtained as nonlinear functions of
the coefficient of normal restitution and the shear rate. These exact
results for the model kinetic equation are also shown to be good
approximations to the corresponding state for the Boltzmann equation via
comparison with direct Monte Carlo simulation for the latter.
\end{abstract}

\draft

\section{Introduction}

\label{sec1}

An idealized representation of rapid flow granular media is given by the
Boltzmann equation for a gas of smooth, hard, inelastic spheres. \cite{C90}
This kinetic equation provides a basis for the study of a wide range of
transport properties, and significant progress has been made in recent years
using numerical methods such as direct Monte Carlo simulation. \cite
{Simulation} In contrast, analytic studies have been limited to homogeneous
states or to those with small spatial gradients. \cite{BDKS98,SG98,GD99}
Even in the homogenous case there is no exact solution corresponding to the
Maxwellian distribution for elastic collisions. Recently, the method of
kinetic models has been proposed for practical access to transport in more
complex states. \cite{BDS99} This method has proven successful for elastic
collisions, where several exact solutions to the model kinetic equations
have been obtained for spatially inhomogeneous states far from equilibrium. 
\cite{Z79,SBG86,BSD87,KDSB89,MG98} Such methods are potentially more
important for granular gases since the states of interest are typically
driven by external boundary conditions, posing intractable difficulties for
solution to the Boltzmann equation. One illustration is the exact solution
of a kinetic model for uniform shear flow, where the rheological properties
were calculated and shown to be in good agreement with those obtained from
simulation of the Boltzmann equation{\ at low density \cite{BRM97} and the
Enskog equation at finite densities. \cite{MGSB99} }{\ Uniform} shear flow
is perhaps the most extensively studied inhomogeneous state for inelastic
particles. \cite{JS83,LSJC84,HS92,SGN96,C00} On the other hand, a more
realistic shearing state is the planar Couette flow, \cite
{RC88,HJR88,L96,B97} where temperature and density gradients coexist with
the velocity field. The objective here is to provide an exact solution to a
model kinetic equation corresponding to Couette flow with {\em arbitrarily\/}
large temperature and flow field gradients, and {\em arbitrary\/}
inelasticity. The results extend a previous exact analysis of momentum and
heat transport far from equilibrium for a gas with elastic collisions. \cite
{BSD87,KDSB89,MG98} Comparison with simulation of the Boltzmann equation for
inelastic collisions again shows good agreement, confirming that the kinetic
model is not only instructive but practical as well.

{\ As noted above, exact solutions in kinetic theory for spatially
inhomogeneous states are exceedingly rare. When, furthermore, such a
solution corresponds to a hydrodynamic state far from equilibrium a unique
and important benchmark is obtained for both conceptual and computational
issues. In the case of granular gases there are two interesting examples:
uniform shear flow and Couette flow. Both allow controlled discussion of
nonlinear rheological properties that are important for a wide class of real
granular flows. Uniform shear flow is a useful idealization, but the
combined heat and momentum transport of Couette flow {\ considered here} is
more realistic and a stronger test of hydrodynamic transport. {\ For uniform
shear, the condition of stationarity imposes a relationship between the
shear rate and the coefficient of restitution, at fixed temperature.}
However, for Couette flow the shear rate and the coefficient of restitution
are independent variables since the temperature profile is allowed to
change. The parameter space for testing hydrodynamics is now two
dimensional, with uniform shear flow recovered as a limiting line
corresponding to zero curvature for the temperature profile. This is
discussed further below. }

In light of some earlier speculation that a hydrodynamic description for
granular flows might be limited to weak dissipation and/or weakly
inhomogeneous states, the results {\ of this paper} {\ provide an example to
the contrary: a) a hydrodynamic description applies, since all space
dependence of the heat and momentum fluxes occurs via explicit functionals
of the hydrodynamic fields and the macroscopic balance equations become a
closed set of hydrodynamic equations which determine the non-trivial space
dependence of the hydrodynamic fields; b) the hydrodynamic description
applies even at strong dissipation and strong inhomogeneity (i.e., beyond
the Navier-Stokes limit). The exact heat and momentum fluxes are
characterized by generalizations of Fourier's law and Newton's viscosity
where the thermal conductivity and viscosity are functions of the shear
rate. The viscometric functions (normal stresses) are spatially constant,
but non-trivial functions of the shear rate as well.}

These strong results are obtained in the context of a simplified model of
the Boltzmann equation. The mathematical and physical basis for this model
as a good representation of the Boltzmann equation is discussed in Ref.\ 
\onlinecite{BDS99}. However, application of the model far from equilibrium
as done here raises the question of its limitations,{\ both quantitative and
qualitative}. The Monte Carlo simulations of the Boltzmann equation {\
provide a more secure basis for the study of \ this macroscopic state.} The
good agreement obtained in this paper for this complex hydrodynamic state
extends that {\ previously demonstrated} for uniform shear flow,\cite
{BRM97,MGSB99} providing additional support for future applications of the
kinetic model to address other realistic boundary driven problems {\ (e.g.,
vibrated columns or vertical flow through a chute in a gravitational field)}. {\ 
Additional concerns are associated with the stability of the stationary
state obtained here, particularly for large spatial gradients and strong
dissipation. No analysis of this problem is provided here (no indication of
instability is seen in the Monte Carlo simulation, although the imposed
planar symmetry may suppress some possible mechanisms).}

{\ This paper is organized as follows.} {\ In the next Section, the state of
macroscopic Couette flow is first considered at the level of the balance
equations for mass, energy, and momentum. Based on previous analysis of this
problem for elastic collisions, it is postulated that the balance equations
support a solution with constant pressure and constant (dimensionless) shear
rate. For consistency, it is shown that the heat and momentum fluxes must be
given by generalizations of Fourier's heat law and Newton's viscosity law,
as noted above. The temperature and velocity profiles are then determined in
terms of the coefficients in these fluxes. To confirm that this
self-consistent ansatz for the fluxes is correct, and to determine the
explicit forms for the transport coefficients, an exact solution to the
kinetic equation for this macroscopic state is constructed in Section \ref
{sec2}. Only the main results are quoted in this Section, while the detailed
analysis is provided in the Appendices. } {\ The primary results of this
analysis are summarized as follows: The heat flux is given by a generalized
version of Fourier's law which is linear in the temperature gradient without
any restriction on this gradient being small. The proportionality constant
is a nonlinear function of the shear rate, including anisotropy effects
inducing a heat flux in the direction of flow as well as in the direction of
the temperature gradient. The momentum flux components are characterized by
three scalar functions of the shear rate, a shear viscosity and two
viscometric functions. All of these properties are applicable even far from
the reference homogeneous state and for all values of the shear rate. The
only restriction is to conditions such that viscous heating dominates
collisional cooling so that the temperature in the bulk is higher than that
at the walls. These exact results for the kinetic model are compared with
direct Monte Carlo simulation results for the Boltzmann equation in Section 
\ref{sec5}. The agreement is found to be good. Finally, the results are
summarized and discussed in Section \ref{sec6}.}

\section{Hydrodynamic description}

\label{sec2.0} Consider a low density granular gas of smooth hard spheres ($
d=3$) or disks ($d=2$) of diameter $\sigma $ and mass $m$. Collisions
between particles are characterized through a constant coefficient of normal
restitution $\alpha $ with values $0<\alpha \leq 1$, the largest value
corresponding to the elastic limit. The macroscopic balance equations for
mass, energy, and momentum are 
\begin{equation}
D_{t}n+n\nabla \cdot {\bf u}=0,  \label{2.6}
\end{equation}
\begin{equation}
D_{t}T+\frac{2}{dn}\left( P_{ij}\nabla _{j}u_{i}+\nabla \cdot {\bf q}\right)
=-\zeta T,  \label{2.7}
\end{equation}
\begin{equation}
D_{t}u_{i}+(mn)^{-1}\nabla _{j}P_{ij}=0,  \label{2.8}
\end{equation}
where $n$ is the density, $T$ is the granular temperature, ${\bf u}$ is the
flow velocity, and $D_{t}=\partial _{t}+{\bf u}\cdot \nabla $ is the
material derivative. In addition, $\zeta $ is the cooling rate (related to
the collisional energy dissipation), ${\sf P}$ is the pressure tensor
(related to the transport of momentum), and ${\bf q}$ is the heat flux
(related to the transport of energy). Whenever these fluxes can be expressed
in terms of the hydrodynamic fields, Eqs.\ (\ref{2.6})--(\ref{2.8}) become a
closed set of hydrodynamic equations for these fields.

\begin{figure}[tbp]
\begin{center}
\parbox{0.45\textwidth}{
\epsfxsize=\hsize \epsfbox{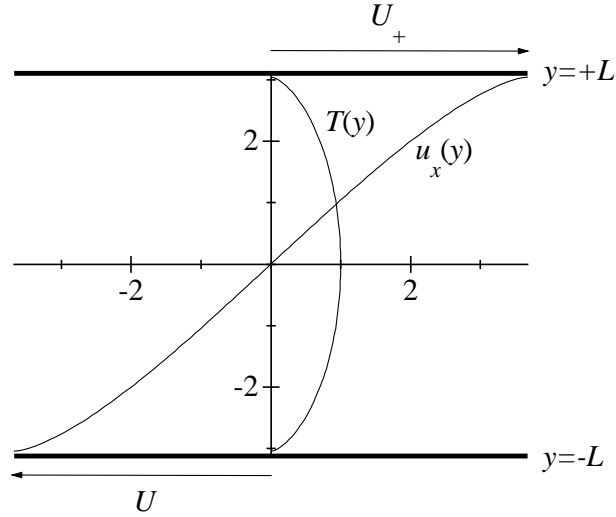}}
\caption{{ Sketch of the geometry of the system. The velocity and
temperature profiles correspond to the solution of the kinetic model for a
coefficient of restitution $\alpha=0.8$ and a reduced shear rate $
a^*=0.95$ [cf.\ Eqs.\ (\protect\ref{n16}) and (\protect\ref{n18})]. The units on 
the scale
are such that $m=1$, $T_0=1$, and $\eta_0(T_0)/p=1$, where $T_0$ is
the maximum temperature across the system.}}
\label{fig0}
\end{center}
\end{figure}The specific problem considered in this paper is steady Couette flow 
{ 
[cf.\ Fig.\ \ref{fig0}]}. The gas of inelastic hard spheres is enclosed
between two parallel plates at $y=\pm L$ in relative motion along the $x$-axis 
and maintained, { in general}, at different temperatures.{ The
resulting flow velocity is along the }$x${ axis and, {}from symmetry,
it is expected that the hydrodynamic fields vary only in the }$y${
direction}. The pressure tensor and the heat flux characterize a macroscopic
state with combined heat and momentum transport. The objective here is to
determine the hydrodynamic fields. Under the above conditions, Eq. (\ref{2.6}) 
is identically satisfied, while the balance equations (\ref{2.7}) and (\ref
{2.8}) yield 
\begin{equation}
\frac{2}{dn}\left( P_{xy}\frac{\partial u_{x}}{\partial y}+\frac{\partial
q_{y}}{\partial y}\right) =-\zeta T,  \label{n4}
\end{equation}
\begin{equation}
\frac{\partial P_{xy}}{\partial y}=0,  \label{n5}
\end{equation}
\begin{equation}
\frac{\partial P_{yy}}{\partial y}=0.  \label{n6}
\end{equation}
Equations (\ref{n4})--(\ref{n6}) are still exact. The temperature $T$ is
defined in the usual way \ in terms of the average kinetic energy, and is
related to the pressure $p$ through the ideal gas law 
\begin{equation}
p=nT=\frac{1}{d}P_{ii}.  \label{n7}
\end{equation}
The second equality follows from the definition of the pressure tensor (see
next Section). There are three independent hydrodynamic fields, which are
taken here to be the \ pressure, temperature, and the $x$ component of the
flow velocity (the other $d-1$ components vanish). The boundary conditions
impose a global shear rate given by $\left[ u_{x}\left( L\right)
-u_{x}\left( -L\right) \right] /2L$. It is useful to introduce a {\em 
dimensionless local\/} shear rate $a$ 
\begin{equation}
a=\frac{1}{\nu _{0}}\frac{\partial u_{x}}{\partial y}.  \label{n8}
\end{equation}
Here $\nu _{0}$ is a convenient collision frequency $\nu _{0}=p/\eta _{0}$, 
\cite{note2} where $\eta _{0}$ is the Navier-Stokes shear viscosity in the 
{\em elastic\/} limit. The variable $a$ will be one of the control
parameters for the Couette flow state. It has the simple physical
interpretation as the ratio of the mean free path $\ell =v_{0}/\nu _{0}$
(where $v_{0}=\sqrt{2T/m}$ is the thermal velocity) to the relevant
hydrodynamic length $h=v_{0}(\partial u_{x}/\partial y)^{-1}$. We now look
for special solutions to the \ macroscopic balance equations characterized
by constant pressure and constant $a$, 
\begin{equation}
p=\text{constant},\quad a=\text{constant}.  \label{n9}
\end{equation}

To provide a closed set of equations for the hydrodynamic fields, the heat
and momentum fluxes must be expressed in terms of these fields. { Forms
for these fluxes are postulated in this Section, as a generalization of
those obtained from a similar analysis of Couette flow for a fluid with
elastic collisions.\cite{BSD87,KDSB89,MG98} Their verification is given in
the following Section and in Appendix \ref{appC}. }Momentum transport is
typically characterized by three scalar rheological functions, the shear
viscosity and two viscometric functions. The $xy$ element for shear stresses
is represented in the form of a generalized Newton's viscosity law 
\begin{equation}
P_{xy}=-F_{\eta }({a},\alpha )\eta _{0}\frac{\partial u_{x}}{\partial y}.
\label{4.16}
\end{equation}
The function $F_{\eta }({a},\alpha )\eta _{0}$ is the generalized shear
viscosity, where the dissipation and all nonlinear rheological effects are
included through the factor $F_{\eta }({a},\alpha )$. The normal stress
components are specified in terms of the dimensionless viscometric functions 
\begin{equation}
{\Psi }_{1}(a,\alpha )=\frac{P_{yy}-P_{xx}}{p{a}^{2}},\quad {\Psi }_{2}(a,\alpha 
)=\frac{P_{zz}-P_{yy}}{p{a}^{2}}.  \label{4.17}
\end{equation}
The heat flux is represented as 
\begin{equation}
q_{i}=-\lambda _{ij}\nabla _{j}T.  \label{4.18}
\end{equation}
The tensor $\lambda _{ij}$ depends on two independent scalar ``transport
coefficients'' 
\begin{equation}
\lambda _{ij}=\lambda _{0}\left[ F_{\lambda }({a},\alpha )\delta _{ij}+\Phi
\left( {a},\alpha \right) {a}_{ij}\right] ,  \label{4.19}
\end{equation}
where $a_{ij}=a\delta _{ix}\delta _{jy}$. The first term of Eq.\ (\ref{4.19}) 
yields a generalization of Fourier's law with the thermal conductivity $
\lambda _{0}F_{\lambda }({a},\alpha )$ modified by the nonlinear rheological
factor $F_{\lambda }({a},\alpha )$. The second term of (\ref{4.19}) provides
information on the anisotropy created by the Couette flow since it gives a
heat flux along the $x$-axis due to a thermal gradient parallel to the $y$-axis. 
This effect is also nonlinear with no analogue at Navier-Stokes
order, being first order in both the shear rate and the temperature
gradient. { In fact, Eqs.\ (\ref{4.18}) and (\ref{4.19}) are a
generalization of the heat flux constitutive equation to Burnett order \cite
{CC70} for the geometry of the problem in the elastic case, according to
which $\Phi (a=0,\alpha =1)=-(\theta _{5}-\theta _{4}\partial \ln \eta
_{0}/\partial \ln T)\eta _{0}/2m\lambda _{0}=-3.5 \text{ (Boltzmann equation)
or }-2.8 \text{ (BGK model)}$}. 

Since $p$ and $a$ are constant, Eqs.\ (\ref{4.16}) and (\ref{4.17}) are
consistent with Eqs.\ (\ref{n5}) and (\ref{n6}). Dimensional analysis shows
that 
\begin{equation}
\eta _{0}\propto T^{1/2},\quad \lambda _{0}\propto T^{1/2},\quad \zeta
\propto T^{-1/2},  \label{4.20}
\end{equation}
where it is understood that the independent variables are $p$ and $T$. Then
Eq.\ (\ref{n4}) becomes 
\begin{equation}
T^{1/2}\frac{\partial }{\partial y}T^{1/2}\frac{\partial T}{\partial y}=-
\frac{2mTp^{2}}{\eta _{0}^{2}}\text{Pr}\gamma ,  \label{4.21}
\end{equation}
where $\gamma \left( a,\alpha \right) $ is the dimensionless constant
defined by 
\begin{equation}
\text{Pr}\gamma =\frac{\eta _{0}}{2m\lambda _{0}F_{\lambda }}\left(
a^{2}F_{\eta }-\frac{d}{2p}\eta _{0}\zeta \right)   \label{4.22}
\end{equation}
and $\text{Pr}=(d+2)\eta _{0}/2m\lambda _{0}$ is the Prandtl number.
Equation (\ref{4.21}) is a closed equation that determines $T=T({a},\alpha
,y)$. With this known, the velocity field is determined from the condition $
a=$ constant 
\begin{equation}
\frac{\partial u_{x}}{\partial y}=a\frac{p}{\eta _{0}(T)}.  \label{4.23}
\end{equation}
{} Before giving the solution, it is \ instructive to express the
temperature in terms of the flow velocity: $T(y)\rightarrow T(u_{x})$. Then
Eqs.\ (\ref{4.21}) and (\ref{4.23}) give directly 
\begin{equation}
\frac{\partial ^{2}T}{{\partial }u_{x}^{2}}=-\frac{2m}{a^{2}}\text{Pr}\gamma
.  \label{n14}
\end{equation}
The sign of the constant $\gamma $ is a result of the competition between
viscous heating and inelastic cooling (represented by the first and second
terms, respectively, on the right-hand side of (\ref{4.22})). If the
dissipation is sufficiently low and/or the shear rate is sufficiently large
so that $\zeta <(2pF_{\eta }/d\eta _{0})a^{2}$, then $\gamma >0$ and the
temperature profile is convex. Equation (\ref{n14}) implies that the
temperature is simply a quadratic function of the flow velocity, 
\begin{equation}
T=T_{0}\left[ 1-c^{2}(u_{x}-u_{0})^{2}\right] ,\quad c^{2}\equiv \frac{m}{
T_{0}a^{2}}\text{Pr}\gamma ,  \label{n16}
\end{equation}
where { $u_{0}$ is an arbitrary constant and $T_{0}$ is the temperature
at the point where $\partial T/\partial u_{x}=0$}. { Here and below
attention is restricted to the case  }$\gamma >0${ which implies that
the shear rate must be larger than a certain critical value, }$
a>a_{c}(\alpha )$. \ The relationship of $T$ \ and $u_{x}$ can be viewed as
a nonequilibrium ``equation of state.'' The explicit $y$-dependence of $u_{x}
$ and $T$ can be easily obtained from Eqs.\ (\ref{n16}) and (\ref{4.23}),
where the constants are determined by the boundary conditions. For instance,
suppose that $y=y_{0}$ is the point where the temperature presents an
extremum and that, without loss of generality, $u_{x}(y_{0})\equiv u_{0}=0$.
Each particular situation is then characterized by the uniform pressure $p$,
the local temperature $T(y_{0})=T_{0}$ and the shear rate $a$. Equation (\ref
{4.23}) becomes 
\begin{equation}
\left( 1-c^{2}u_{x}^{2}\right) ^{1/2}\frac{\partial u_{x}}{\partial y}=a
\frac{p}{\eta _{0}(T_{0})},  \label{n17}
\end{equation}
whose (implicit) solution is 
\begin{equation}
y=y_{0}+\frac{\eta _{0}(T_{0})}{2cpa}\left[ \sin ^{-1}\left( cu_{x}\right)
+cu_{x}\left( 1-c^{2}u_{x}^{2}\right) ^{1/2}\right] .  \label{n18}
\end{equation}
{ Note that Eqs.\ (\ref{n16}) and (\ref{n18}) are only meaningful for $
u_{x}^{2}\leq c^{-2}$, which implies $|y-y_{0}|\leq \pi \eta _{0}(T_{0})/4cpa
$. The velocity profile (\ref{n18}) and the temperature profile (\ref{n16})
are shown in Fig.\ \ref{fig0} for a representative case ($\alpha =0.8$, $
a=0.95$) with $y_{0}=0$. The actual shear rate $\partial u_{x}/\partial y$
is practically constant across most of the system, but it rapidly increases
as the temperature becomes much smaller than $T_{0}$ [cf.\ Eq.\ (\ref{4.23})].}

To summarize, the conditions of constant pressure and shear rate, Eq. (\ref
{n9}), together with the constitutive relations for the heat and momentum
fluxes, Eqs. (\ref{4.16})--(\ref{4.19}) allow an exact stationary solution
to the macroscopic balance equations. In the next Section the assumed
conditions and constitutive relations are confirmed by an exact solution to
the kinetic equation. In addition, this solution provides explicit
expressions for the transport coefficients $F_{\eta }$, $\Psi _{1,2}$, $
F_{\lambda }$, and $\Phi $ as functions of $a$ and $\alpha $.

\section{Kinetic equation and exact solution}

\label{sec2}

Consider a low density granular gas of smooth hard spheres in $d$
dimensions. At sufficiently low density the phase space distribution
function, $f({\bf r},{\bf v},t)$, is determined from the Boltzmann kinetic
equation modified appropriately for inelastic collisions. Due to the
mathematical complexity of this equation, analysis has been limited to
perturbative approximations for small spatial inhomogeneities. \cite
{BDKS98,SG98,GD99} To describe more general nonequilibrium states it is
useful to replace the Boltzmann equation with a more tractable model kinetic
equation constructed to preserve its most important qualitative features
(e.g., macroscopic balance equations). The model kinetic equation chosen for
analysis here is 
\begin{equation}
\partial _{t}f+{\bf v}\cdot \nabla f=-\nu (f-f_{\ell })+\frac{1}{2}\zeta
\partial _{{\bf v}}\cdot \left[ \left( {\bf v-u}\right) f\right] ,
\label{2.1}
\end{equation}
where $f_{\ell }$ is the local Maxwellian distribution 
\begin{equation}
f_{\ell }({\bf r},{\bf v},t)=n({\bf r},t)\left[ \frac{m}{2\pi T({\bf r},t)}
\right] ^{d/2}\exp \left[ -\frac{m\left( {\bf v}-{\bf u}({\bf r},t)\right)
^{2}}{2T({\bf r},t)}\right] .  \label{2.2}
\end{equation}
This distribution function is parameterized by the nonequilibrium density $n$
, granular temperature $T$, and flow velocity ${\bf u}$, which are defined
in terms of moments of $f$: 
\begin{equation}
n=\int d{\bf v}f,\quad T=\frac{m}{dn}\int d{\bf v}\,V^{2}f,\quad {\bf u}=
\frac{1}{n}\int d{\bf v}\,{\bf v}f,  \label{2.3}
\end{equation}
where ${\bf V=v-u}$ is the velocity relative to the local flow. The right
side of (\ref{2.1}) is a model for the nonlinear Boltzmann collision
operator. The first term describes collisional relaxation towards the local
Maxwellian, with collision rate $\nu $. The second term describes the
dominant collisional cooling effects, where $\zeta $ is the cooling rate. 
{ The necessity for this term to accurately represent the spectrum of the
Boltzmann collision operator is discussed in Ref.\ \onlinecite{BDS99}. However, 
it can be
viewed more simply as an effective ``drag'' force that produces the same
energy loss rate as that produced by the inelastic collisions}. The
parameters $\nu $ and $\zeta $ are chosen for good quantitative agreement of
the viscosity and cooling rate with those obtained from the Boltzmann
equation, 
\begin{equation}
\zeta =\frac{2\pi ^{(d-1)/2}}{d\Gamma (d/2)}\sigma ^{d-1}n\left( \frac{T}{m}
\right) ^{1/2}(1-\alpha ^{2}),\quad \nu =\nu _{0}-\zeta ,  \label{2.4}
\end{equation}
\begin{equation}
\nu _{0}\equiv k\ n\sigma ^{d-1}\left( \frac{\pi T}{m}\right) ^{1/2},
\label{2.4bis}
\end{equation}
where $k=15/16$ for $d=3$ (spheres) and $k=2$ for $d=2$ (disks). Both $\nu $
and $\zeta $ depend on the density and temperature, whose space and time
dependence has been left implicit. Further details motivating these choices
can be found in Ref.\ \onlinecite{BDS99}, where a somewhat more
sophisticated model is described.\cite{note} Equations (\ref{2.1})--(\ref
{2.4}) define the kinetic equation to be applied in this work.

By taking moments with respect to $1$, ${\bf v}$, and $v^{2}$, this model
kinetic equation yields the same form of the macroscopic balance equations
for mass, energy, and momentum, Eqs.\ (\ref{2.6})--(\ref{2.8}), as those
given from the Boltzmann equation, thus confirming the interpretation of $
\zeta $ as the cooling rate. The pressure tensor ${\sf P}$ and the heat flux 
${\bf q}$ are given by 
\begin{equation}
{\sf P}=m\int d{\bf v}\,{\bf V} {\bf V} f\text{ },\quad {\bf q}= \frac{m}{2}
\int d{\bf v}\,V^{2} {\bf V} f .  \label{2.9}
\end{equation}

We now consider the specific problem of steady Couette flow described in the
previous Section. The main objective is to verify the constitutive equations
(\ref{4.16})--(\ref{4.19}) from the fundamental definitions (\ref{2.9}), and
to verify the assumed constancy for the pressure and shear rate. For the
chosen geometry the kinetic equation becomes 
\begin{equation}
\left( 1-\zeta ^{\ast }+v_{y}\nu _{0}^{-1}\partial _{y}\right) f-\frac{1}{2}{
\zeta ^{\ast }}\partial _{{\bf v}}\cdot \left( {\bf V}f\right) =(1-\zeta
^{\ast })f_{\ell },  \label{2.14}
\end{equation}
where 
\begin{equation}
\zeta ^{\ast }=\frac{\zeta }{\nu _{0}}=\frac{2\pi ^{(d-2)/2}}{kd\Gamma (d/2)}
(1-\alpha ^{2})  \label{4.13}
\end{equation}
is a constant that henceforth will characterize the inelasticity dependence.
Equation (\ref{2.14}) allows one to understand the structural simplification
afforded by the kinetic model relative to the Boltzmann equation. Assuming
the hydrodynamic fields of Section \ref{sec2.0}, the local equilibrium
distribution function $f_{\ell }$ is completely specified and Eq.\ (\ref
{2.14}) becomes a linear inhomogeneous partial first order differential
equation that is readily solved with specified boundary conditions. This
determines explicitly the entire velocity dependence. However, the solution
is only formal since the assumed fields must satisfy (\ref{2.3}). These are
consistency conditions that are needed to justify the assumed forms for the
hydrodynamic fields. Alternatively, since these fields have been shown to
follow from the constitutive equations (\ref{4.16})--(\ref{4.19}) for the
heat and momentum fluxes in Section \ref{sec2.0}, it is sufficient to show
that these equations are verified by use of this formal solution in (\ref
{2.9}). Both the consistency conditions and the fluxes are verified
explicitly in Appendix \ref{appC}.

The general solution to Eq.\ (\ref{2.14}) is 
\begin{equation}
f(y,{\bf v})=f_{B}(y,{\bf v})+(1-\zeta ^{\ast })\int_{0}^{\infty }dt\
e^{-(1-\zeta ^{\ast }-\frac{d}{2}{\zeta ^{\ast }})t}\ e^{-{\cal D}t}f_{\ell
}(y,{\bf v}),  \label{2.15}
\end{equation}
where ${\cal D}$ is the sum of a generator for translations of $y$ and the
generator for scale transformation of ${\bf V}$, 
\begin{equation}
{\cal D}\equiv v_{y}\nu _{0}^{-1}\partial _{y}-\frac{{\zeta ^{\ast }}}{2}
{\bf V}\cdot \partial _{{\bf v}}.  \label{2.16}
\end{equation}
The explicit effects of this generator are described in Appendix \ref{appA}.
The first term of (\ref{2.15}), $f_{B}(y,{\bf v})$, is a solution to the
homogeneous kinetic equation obtained from (\ref{2.14}) by setting $f_{\ell
}\rightarrow 0$. The detailed form of this contribution is determined by the
chosen boundary conditions. The physical boundary conditions are
specification of the half distributions for velocities directed away from
the walls at $y=\pm L$, given explicitly or implicitly in terms of the
distributions for velocities directed at the walls. The specific
relationship characterizes the motion and temperature of the walls. One
possibility for Couette flow is diffuse conditions, where the distributions
for velocities away from the walls are given by a Maxwellian whose
parameters specify the temperatures $T_{\pm }$ and velocities $U_{\pm }$ of
the walls at $y=\pm L$.

In general $f_{B}(y,{\bf v})$ has a detailed explicit dependence on the
geometry of the system and is responsible for the ``boundary layer'' near
the wall. In contrast, the second term of (\ref{2.15}) is an example of a
``normal'' solution where all of its dependence on $y$ occurs only through
the hydrodynamic fields. Here we consider idealized boundary conditions for
which $f_{B}(y,{\bf v})\rightarrow 0$ so that the entire solution is given
by the second term of (\ref{2.15}). This idealized boundary condition
eliminates boundary layers and admits the possibility of simple hydrodynamic
profiles that are exact throughout the system. Furthermore, the solution to
the kinetic equation is now normal so that the fluxes become functionals of
the hydrodynamic fields and (\ref{2.6})--(\ref{2.8}) become a closed set of
hydrodynamic equations. The idealized boundary conditions correspond to the
limit $T_{\pm }\rightarrow 0$ { (as illustrated in Fig.\ \ref{fig0})}.
This is verified in Appendix \ref{appB} and will not be discussed further
here.

Equation (\ref{2.15}) with these idealized boundary conditions admits an
exact solution for steady Couette flow [see Eq.\ (\ref{c2}) for its explicit
form] characterized by the solutions to Eqs.\ (\ref{n8}), (\ref{n9}), and 
(\ref{4.21}) given in Section \ref{sec2.0}, which are rewritten here for the
sake of completeness: 
\begin{equation}
p(y)=n(y)T(y)=\text{const},\quad {\nu _{0}}^{-1}\partial _{y}u_{x}=a,\quad ({ 
\nu _{0}}^{-1}\partial _{y})^{2}T=-2m\text{Pr}\gamma .  \label{2.18}
\end{equation}
The Prandtl number, defined following (\ref{4.22}), is equal to 1 in the
kinetic model while $\text{Pr}=(d-1)/d$ for the Boltzmann equation. The fact
that $\text{Pr}=1$ in the kinetic model is a consequence of the introduction
of a single collision frequency, that is unable to reproduce simultaneously
the exact shear viscosity and thermal conductivity Navier-Stokes
coefficients. The components of the flow velocity in the $y$ and $z$
directions vanish. The parameter ${a}$ is the dimensionless shear rate and
characterizes the relative velocities of the wall. Together with the
coefficient of restitution $\alpha $ it is a given control parameter in
terms of which all transport properties are expressed. The dimensionless
parameter $\gamma (a,\alpha )$ characterizes the curvature of the
temperature field and is a consequence of both the viscous heating due to
the shear rate and the collisional dissipation characterized by $\alpha $.
The consistency conditions and forms for the fluxes are verified in Appendix 
\ref{appC}, confirming that the distribution function (\ref{2.15}) is an
exact solution to the kinetic equation. This verification also fixes the
functional dependence of ${\gamma }({a},\alpha )$ on the control parameters.
The details of the analysis also are given in Appendix \ref{appC}, yielding
the following implicit equation 
\begin{equation}
1=\frac{4(1-\zeta ^{\ast })}{d\pi ^{1/2}}\int_{0}^{\infty
}du\,e^{-u^{2}}\int_{0}^{\infty }dt\,e^{-t}\left[ \frac{d-1}{2}+\left( 1+{a}
^{2}t^{2}\right) u^{2}\right] \left( 1+2{\gamma }w^{2}\right) ^{-1}.
\label{4.14}
\end{equation}
where 
\begin{equation}
w(u,t)=\frac{2u}{\zeta ^{\ast }}\left( 1-e^{-\frac{1}{2}\zeta ^{\ast
}t}\right) .  \label{4.14a}
\end{equation}
Interestingly, the representation (\ref{4.14}) exists only for ${\gamma }
\geq 0$ or, equivalently, for ${a}$ equal to or larger than some critical
value $a_{c}$ of the shear rate. This critical shear rate corresponds to ${\
\gamma }=0$, or uniform temperature (see Eq. (\ref{n16})). Setting $\gamma =0
$ in Eq.\ (\ref{4.14}), one gets 
\begin{equation}
a_{c}^{2}=\frac{d}{2}\frac{\zeta ^{\ast }}{1-\zeta ^{\ast }}.  \label{4.15}
\end{equation}
For the limiting case $a=a_{c}$ the viscous heating is exactly balanced by
collisional cooling and the gas is in a state of uniform shear flow.\cite
{BRM97} {\ This is further discussed at the end of this Section. }In the
regime of low shear rates and low dissipation, Eq.\ (\ref{4.14}) yields $
\gamma \approx (a^{2}-d\zeta ^{\ast }/2)/(d+2)$ and consequently $
a_{c}^{2}\approx d\zeta ^{\ast }/2$.

The explicit expressions for the momentum and heat fluxes are also derived
in Appendix \ref{appC}. {}From them it is possible to identify the
generalized transport coefficients defined in Eqs.\ (\ref{4.16})--(\ref{4.19}). 
The viscosity coefficient is 
\begin{equation}
F_{\eta }({a},\alpha )=\frac{4(1-\zeta ^{\ast })}{\pi ^{1/2}}
\int_{0}^{\infty }du\,e^{-u^{2}}u^{2}\int_{0}^{\infty }dt\,e^{-t}t\left( 1+2{ 
\gamma }w^{2}\right) ^{-1}.  \label{4.15a}
\end{equation}
The viscometric functions characterizing normal stresses are 
\begin{equation}
{\Psi }_{1}({a},\alpha )=-\frac{2(1-\zeta ^{\ast })}{{a}^{2}\pi ^{1/2}}
\int_{0}^{\infty }du\,e^{-u^{2}}\int_{0}^{\infty }dt\,e^{-t}\left[ 1-2\left(
1-{a}^{2}t^{2}\right) u^{2}\right] \left( 1+2{\gamma }w^{2}\right) ^{-1},
\label{4.18.2}
\end{equation}
\begin{equation}
{\Psi }_{2}({a},\alpha )=\frac{2(1-\zeta ^{\ast })}{{a}^{2}\pi ^{1/2}}
\int_{0}^{\infty }du\,e^{-u^{2}}\left( 1-2u^{2}\right) \int_{0}^{\infty
}dt\,e^{-t}\left( 1+2{\gamma }w^{2}\right) ^{-1}.  \label{4.18.3}
\end{equation}
The thermal conductivity coefficient is given in terms of $\gamma $ and $
F_{\eta }$ by Eq.\ (\ref{4.22}) 
\begin{equation}
F_{\lambda }=\frac{\eta _{0}}{2m\lambda _{0}\text{Pr}\gamma }\left(
a^{2}F_{\eta }-\frac{d}{2p}\eta _{0}\zeta \right).  \label{4.18.3.1}
\end{equation}
Finally, the expression for the cross coefficient $\Phi $ in the heat flux
is 
\begin{eqnarray}
\Phi \left( {a},\alpha \right) &=&-\frac{8(1-\zeta ^{\ast })}{\left(
d+2\right) \pi ^{1/2}}\int_{0}^{\infty }du\,e^{-u^{2}}u\int_{0}^{\infty
}dt\,e^{-\left( 1+\frac{1}{2}{\zeta }^{\ast }\right) t}tw\left[ \frac{d+1}{2}
+\left( 1+a^{2}t^{2}\right) u^{2}\right]  \nonumber \\
&&\times \left( 1+2{\gamma }w^{2}\right) ^{-2}.  \label{4.18.4}
\end{eqnarray}

In summary, the distribution function, hydrodynamic fields, and transport
coefficients for the heat and momentum fluxes have been determined exactly
in terms of the imposed shear rate $a$ and the restitution coefficient $
\alpha $. For small $a$ and $\alpha $ the results agree with predictions of
the Navier-Stokes hydrodynamics. More generally, they extend the description
of Couette flow to large spatial inhomogeneity and strong dissipation. The
final results are still only implicit, but the entire problem has been
brought to quadratures. A numerical evaluation of these expressions is
provided in the next Section for comparison with numerical simulation of the
Boltzmann kinetic equation.

It is easy to check that all results presented in this Section reduce to
those previously derived in the elastic limit $\alpha =1$ by using the
Bhatnagar-Gross-Krook (BGK) model. \cite{BSD87,KDSB89,MG98} {\ In addition,
the results of this Section include as a limiting case those corresponding
to the uniform shear flow for an inelastic gas. \cite{BRM97,MGSB99} This
happens when the shear rate and the inelasticity combine to yield a zero
curvature for the temperature profile ($\gamma \rightarrow 0$). \ Equation 
(\ref{4.15}) can be seen as the relationship between the reduced shear rate
and the coefficient of restitution, which are not independent in the uniform
shear flow problem. The transport coefficients in this limiting case are
given in Appendix \ref{appD}. It is also interesting to note that this
uniform shear flow state is mathematically equivalent to the corresponding
case of elastic collisions with an external thermostat force $-\case{1}{2}
m\zeta {\bf V}$ adjusted to control the viscous heating. \cite{USF} In this
context, Eq.\ (\ref{4.15}) (with the adequate change of units) gives the
shear rate dependence of the thermostat and this coincides with the previous
result derived in the uniform shear flow problem {\ for an elastic gas}, as
expected.}

\section{Comparison with Monte Carlo simulations}

\label{sec5} In this Section we compare the predictions of the kinetic model
with computer simulations of the Boltzmann equation by means of the Direct
Simulation Monte Carlo (DSMC) method. Although this method was originally
devised for elastic particles, \cite{Bird} its extension to the inelastic
case is straightforward. \cite{Simulation,MGSB99} This method has proven to
be an efficient and reliable tool for solving numerically the Boltzmann
equation.

We have used the adaptation of the DSMC method to the steady Couette flow
for a granular gas following the same steps as those recently worked out in
the elastic case, \cite{MGS00} so that the technical details of the
simulations are omitted here. We have considered a system of hard spheres ($
d=3$) with three different values for the coefficient of restitution: $
\alpha=1$ (elastic case), $\alpha=0.9$, and $\alpha=0.8$. For each case,
typically four or five different shearing states have been taken. Once a
steady state is reached, the profiles of the hydrodynamic quantities and of
the fluxes are measured in the bulk region. {}From these profiles the
transport coefficients are identified.

\begin{figure}[tbp]
\begin{center}
\parbox{0.45\textwidth}{
\epsfxsize=\hsize \epsfbox{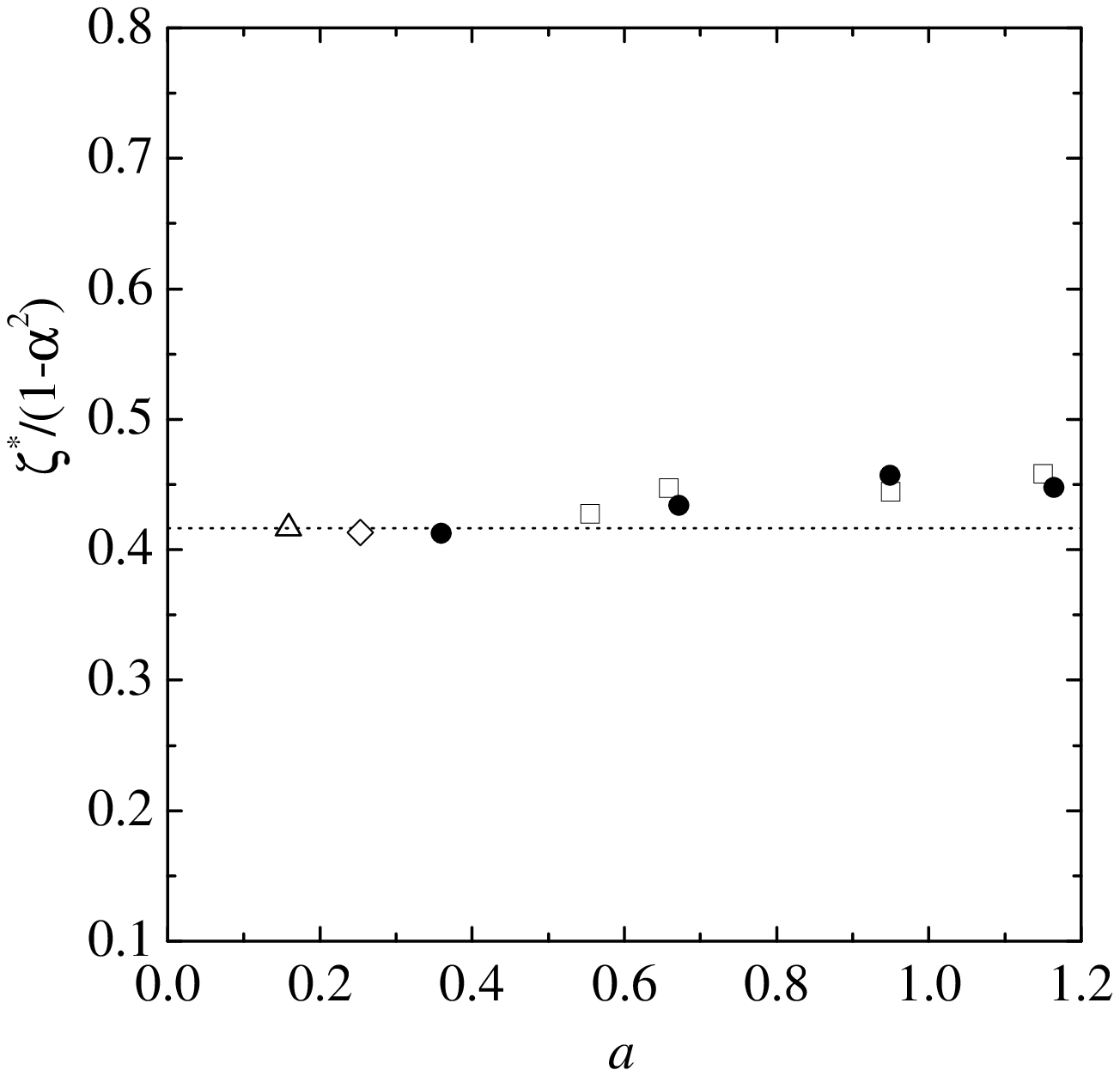}}
\caption{ Plot of the reduced cooling rate $\zeta ^{\ast }$ divided
by $1-\alpha ^{2}$ versus the reduced shear rate $a$, as obtained
from simulation for $\alpha =0.98$ (triangle), 0.95 (diamond), 0.9
(circles), and 0.8 (squares). The dotted line represents the local
equilibrium approximation $\zeta ^{\ast }/(1-\alpha 
^{2})=5/12$ used in the kinetic model. }
\label{fig1}
\end{center}
\end{figure}

\begin{figure}[tbp]
\begin{center}
\parbox{0.45\textwidth}{
\epsfxsize=\hsize \epsfbox{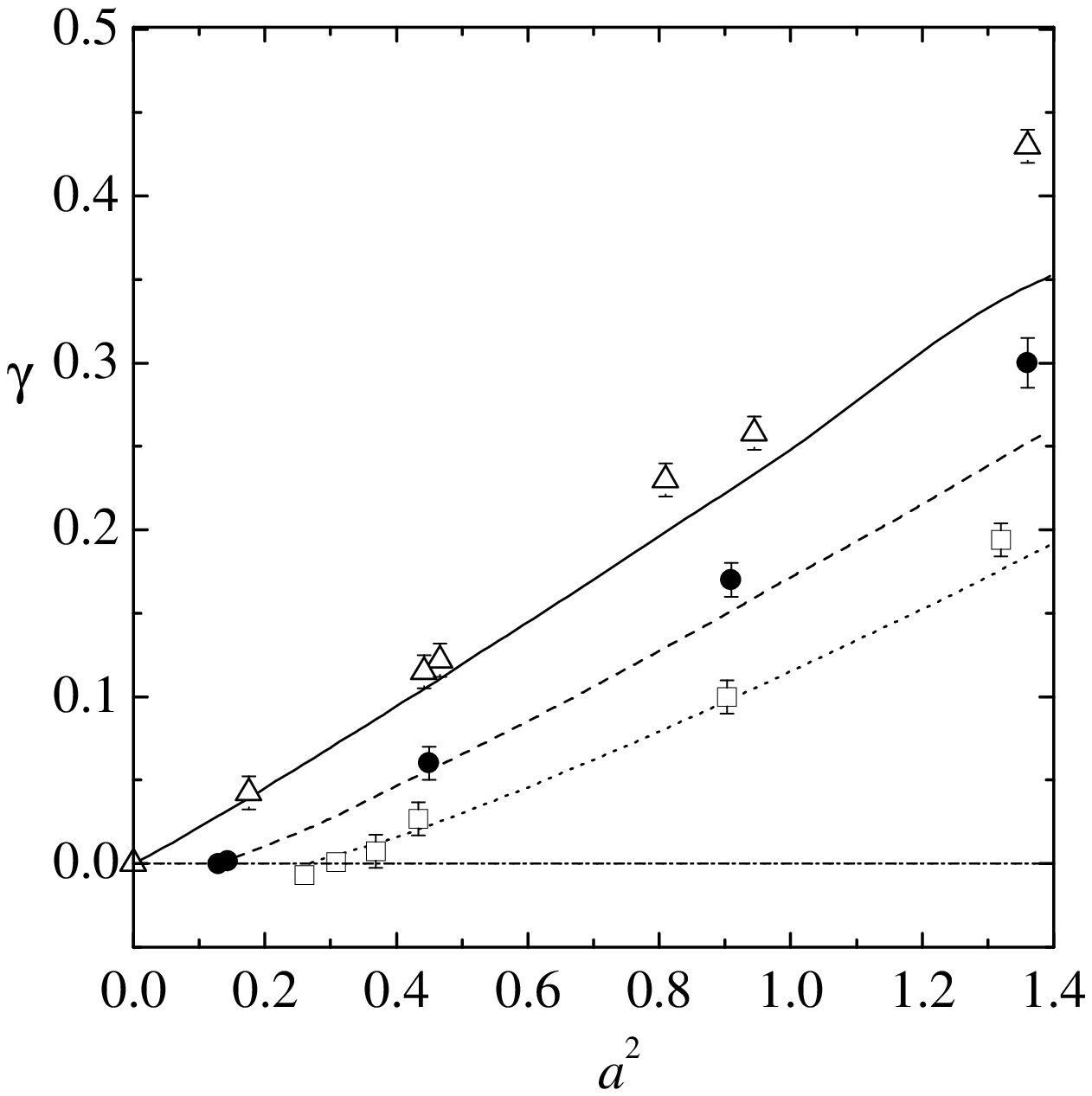}}
\caption{ Plot of the thermal curvature parameter $\gamma $ as a
function of $a^{2}$, as obtained from the kinetic model (lines) and from
simulation (symbols), for $\alpha =1$ (solid line and triangles),
0.9 (dashed line and circles), and 0.8 (dotted line and squares). }
\label{fig2}
\end{center}
\end{figure}

\begin{figure}[tbp]
\begin{center}
\parbox{0.45\textwidth}{
\epsfxsize=\hsize \epsfbox{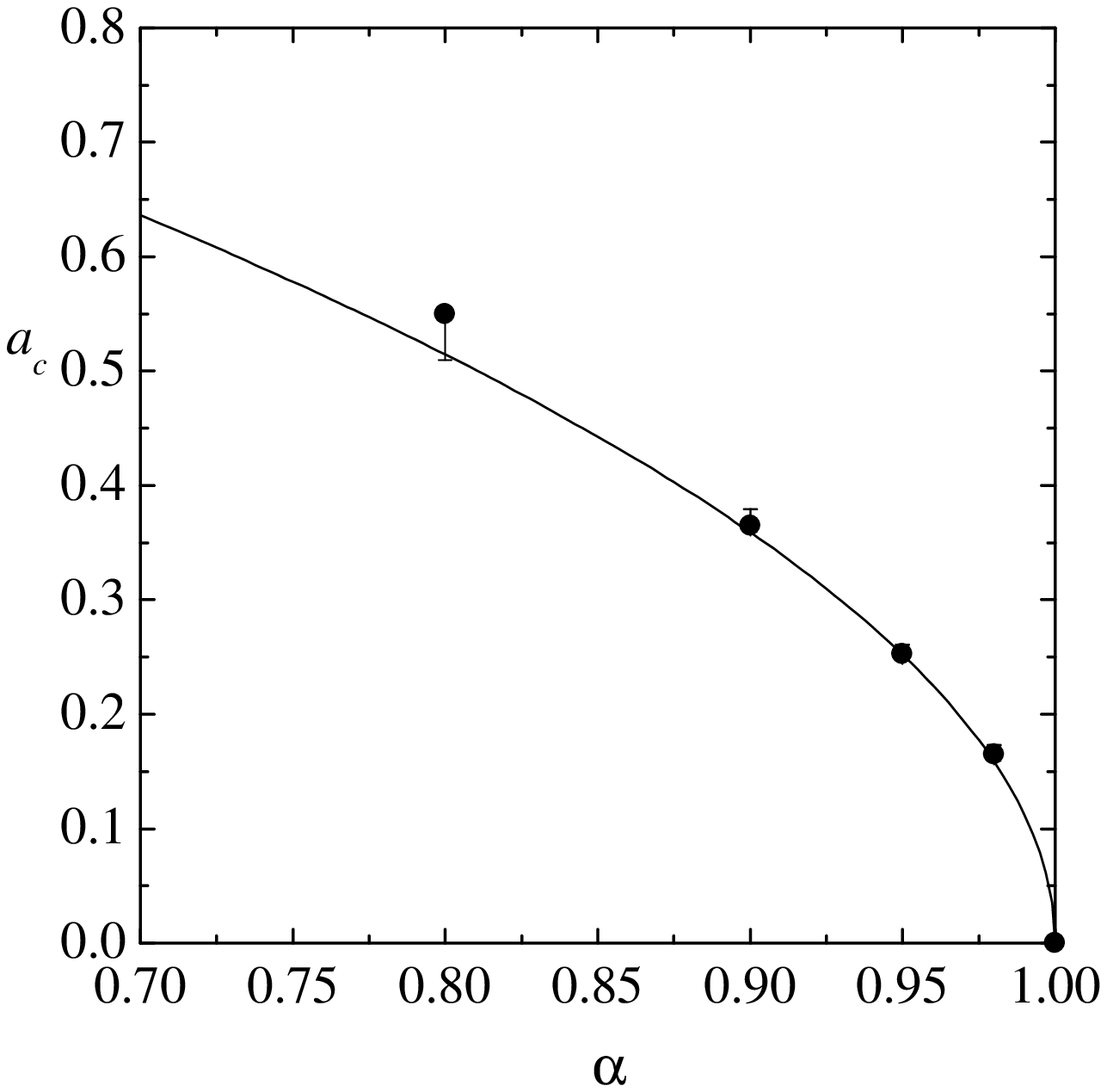}}
\caption{ Plot of the critical shear rate $a_{c}$ as a function of the
coefficient of restitution $\alpha $, as obtained from the kinetic
model (solid line) and from simulation (circles). }
\label{fig3}
\end{center}
\end{figure}

The two parameters defining the kinetic model are the collision frequency $
\nu $ and the cooling rate $\zeta $, given by (\ref{2.4}). These choices
were made to provide a quantitative representation of the shear viscosity at
the Navier-Stokes level, while retaining the relative simplicity of the
kinetic equation. To verify the extent to which these choices are valid for
larger spatial gradients we compare the cooling rate of the kinetic model
with that obtained from the Boltzmann equation. Both are proportional to $
(1-\alpha ^{2})$. In Fig.\ \ref{fig1} we plot the ratio $\zeta ^{\ast
}/(1-\alpha ^{2})$ obtained from the simulation versus the shear rate for
different values of $\alpha $. It is apparent that $\zeta ^{\ast }$ is well
approximated by the kinetic model form, although the latter tends to
underestimate the correct value, especially as the shear rate increases.
This supports the expectation that the kinetic model retains quantitative as
well as qualitative validity even for the extreme conditions admitted by the
exact solution of the previous Sections.
\vspace{0.5cm}
The simulation results indicate that the pressure $p$ is practically
constant, in agreement with the first assumption in (\ref{2.18}). Also, the
actual shear rate $\partial _{y}u_{x}$ multiplied by $T^{1/2}$ is
practically constant, allowing identification of a reduced shear rate $a$ in
agreement with the second equality in (\ref{2.18}). We have also checked
that the temperature profile is well characterized by the third equation in 
(\ref{2.18}) (with $\text{Pr}=2/3$). The parameter $\gamma $ characterizing
the curvature of the temperature profile is plotted in Fig.\ \ref{fig2} as a
function of the shear rate for $\alpha =0.8$, $0.9$, and $1$. The agreement
between theory and simulation is quite good. Surprisingly, this is true even
for strong inelasticity. There is a remarkable influence of the degree of
dissipation on $\gamma $ at a given value of the shear rate. As the
coefficient of restitution $\alpha $ decreases, so does the temperature
variation across the system, so that the state tends to resemble the uniform
shear flow, in which case the kinetic model is known to be quite accurate. 
\cite{BRM97,MGSB99} Figure \ref{fig2} also confirms that, for a given value
of $\alpha $, there exists a critical shear rate $a_{c}$ such that $\gamma
\rightarrow 0^{+}$ as $a$ tends to $a_{c}$ from above. For shear rates
smaller than the critical value $a_{c}$ the collisional cooling effect
dominates over the viscous heating and the temperature presents a minimum
instead of a maximum, which could be characterized by a negative $\gamma $. 
\cite{BC98,Cordero} However, the solution obtained here does not extend to
negative $\gamma $, as noted above. The dependence of $a_{c}$ on the
coefficient of restitution $\alpha $ is shown in Fig.\ \ref{fig3}. The
simple prediction (\ref{4.15}) reproduces quite well the simulation data. {
Figure \ref{fig3} can also be interpreted as the two dimensional parameter
space of the inelastic Couette flow. The elastic Couette problem corresponds
to the line $\alpha =1$ and arbitrary $a$. On the other hand, the uniform
shear flow for a dilute granular gas is represented by the line $
a=a_{c}(\alpha )$ for arbitrary $\alpha $. The case studied in this paper
applies to the region enclosed between both lines, namely, for arbitrary $
\alpha $ and $a$, provided that $a\geq a_{c}(\alpha )$.}
\vspace{0.5cm}

\vspace{0.5cm}

\begin{figure}[tbp]
\begin{center}
\parbox{0.45\textwidth}{
\epsfxsize=\hsize \epsfbox{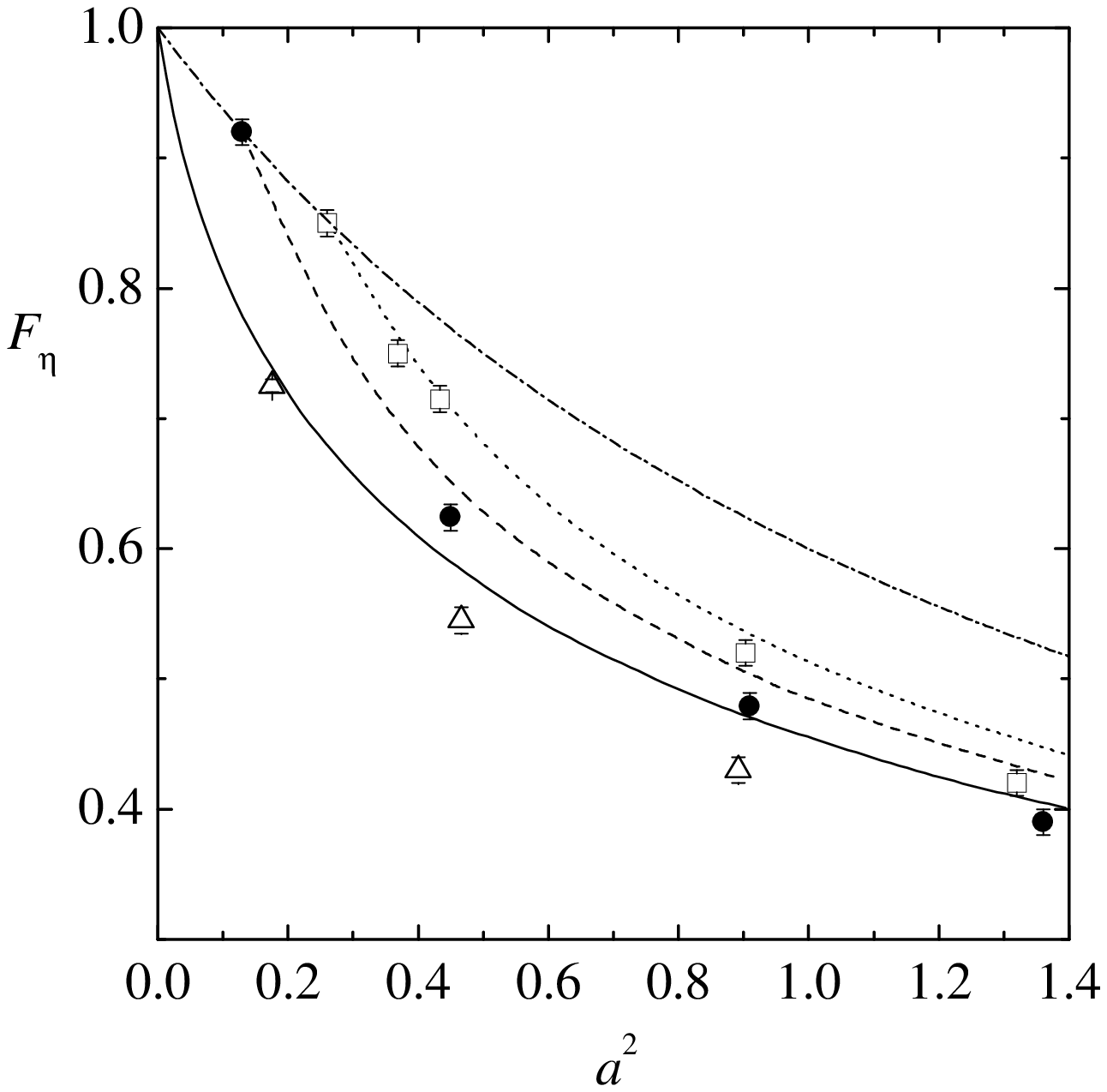}}
\caption{{Plot of the viscosity function $F_\eta $ as a
function of $a^{2}$, as obtained from the kinetic model (lines) and from
simulation (symbols), for $\alpha =1$ (solid line and triangles),
0.9 (dashed line and circles), and 0.8 (dotted line and squares). The
dash-dotted line refers to the corresponding transport coefficient on the
critical line (uniform shear flow).} }
\label{fig4}
\end{center}
\end{figure}

\begin{figure}[tbp]
\begin{center}
\parbox{0.45\textwidth}{
\epsfxsize=\hsize \epsfbox{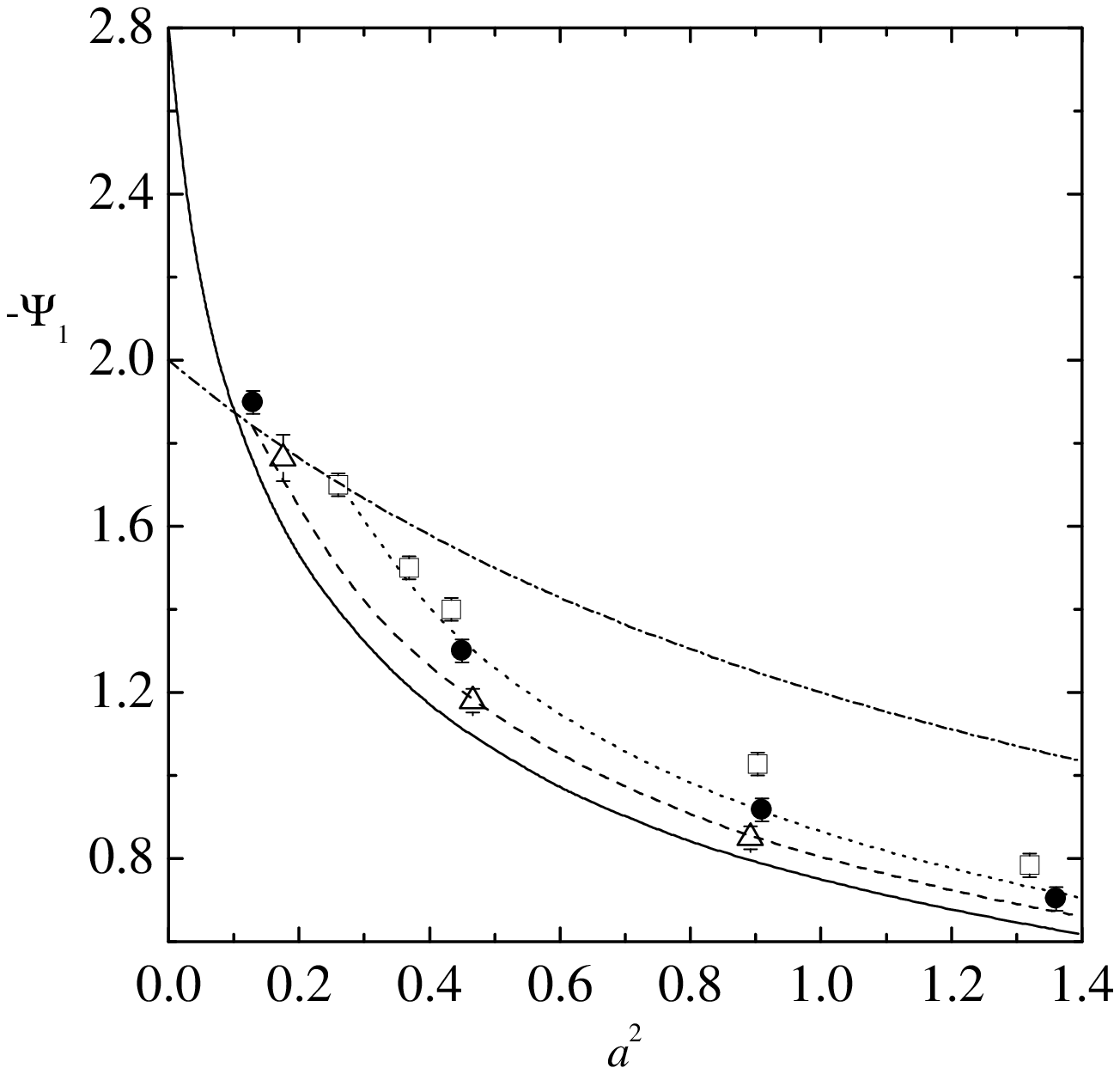}}
\caption{ The same as in Fig.\ \ref{fig4} but for the first viscometric
function $\Psi _{1}$. }
\label{fig5}
\end{center}
\end{figure}

\begin{figure}[tbp]
\begin{center}
\parbox{0.45\textwidth}{
\epsfxsize=\hsize \epsfbox{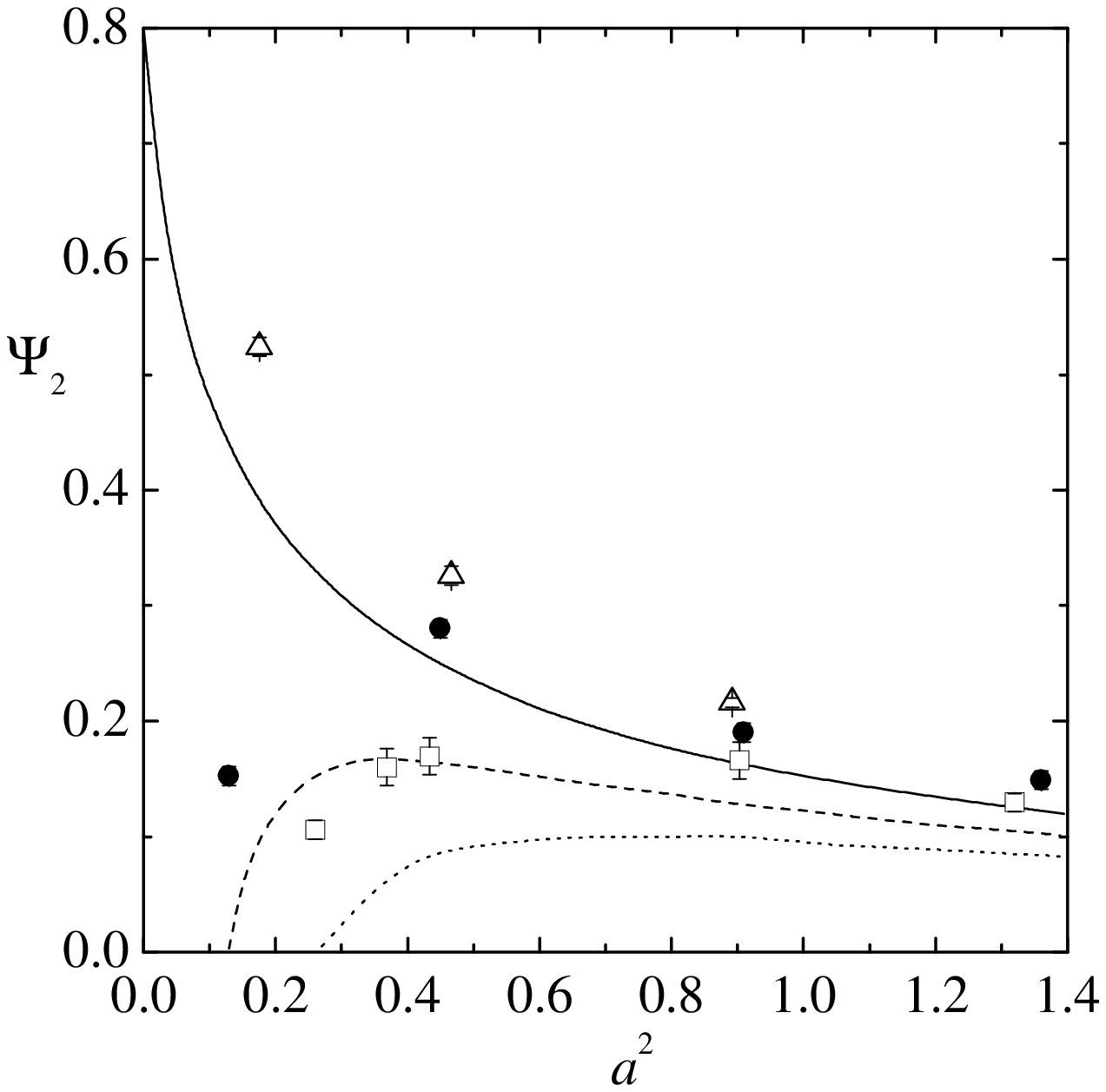}}
\caption{ The same as in Fig.\ \ref{fig4} but for the second viscometric
function $\Psi _{2}$. Note that $\Psi _{2}=0$ on the critical line. }
\label{fig6}
\end{center}
\end{figure}

Now we compare the simulation results and the theoretical predictions for
the five generalized transport coefficients in the domain $a\geq a_{c}$. {\
For the sake of clarity, we have also plotted the curves representing the
transport coefficients for the uniform shear flow when eliminating the
coefficient of restitution in favor of the shear rate.} In Fig.\ \ref{fig4}
we plot the shear rate dependence of the viscosity function $F_{\eta }$
defined by Eq.\ (\ref{4.16}). Needless to say, this is the most relevant
quantity in a shearing state. Regardless of the value of $\alpha $, shear
thinning effects are present, i.e., $F_{\eta }$ decreases as the shear rate
increases. This rheological behavior is quantitatively well described by the
kinetic model, again especially for large dissipation. In general, however,
the model tends to overestimate the value of $F_{\eta }$. The two
viscometric functions (\ref{4.17}) are plotted in Figs.\ \ref{fig5} and \ref
{fig6}. While the agreement between the theoretical predictions and the
simulation results in the case of $\Psi _{1}$ is similar to the one observed
with $F_{\eta }$, the agreement in the case of $\Psi _{2}$ is only
qualitative. In particular, the model succeeds in capturing the
non-monotonic behavior of $\Psi _{2}$ in the inelastic case.
\vspace{0.5cm}

\vspace{0.5cm}

\vspace{0.5cm}

\begin{figure}[tbp]
\begin{center}
\parbox{0.45\textwidth}{
\epsfxsize=\hsize \epsfbox{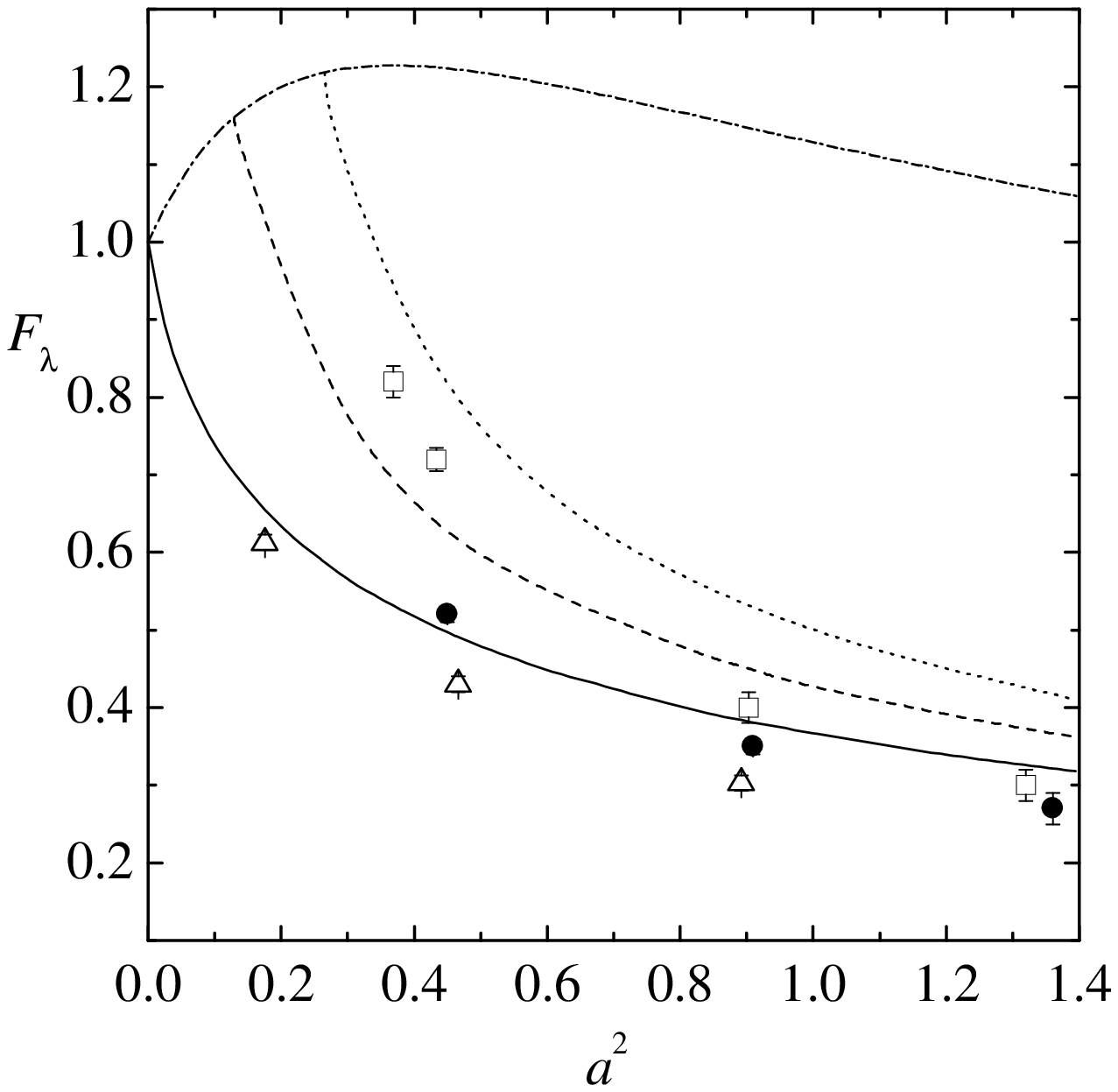}}
\caption{ The same as in Fig.\ \ref{fig4} but for the thermal conductivity
function $F_\lambda $. }
\label{fig7}
\end{center}
\end{figure}

\begin{figure}[tbp]
\begin{center}
\parbox{0.45\textwidth}{
\epsfxsize=\hsize \epsfbox{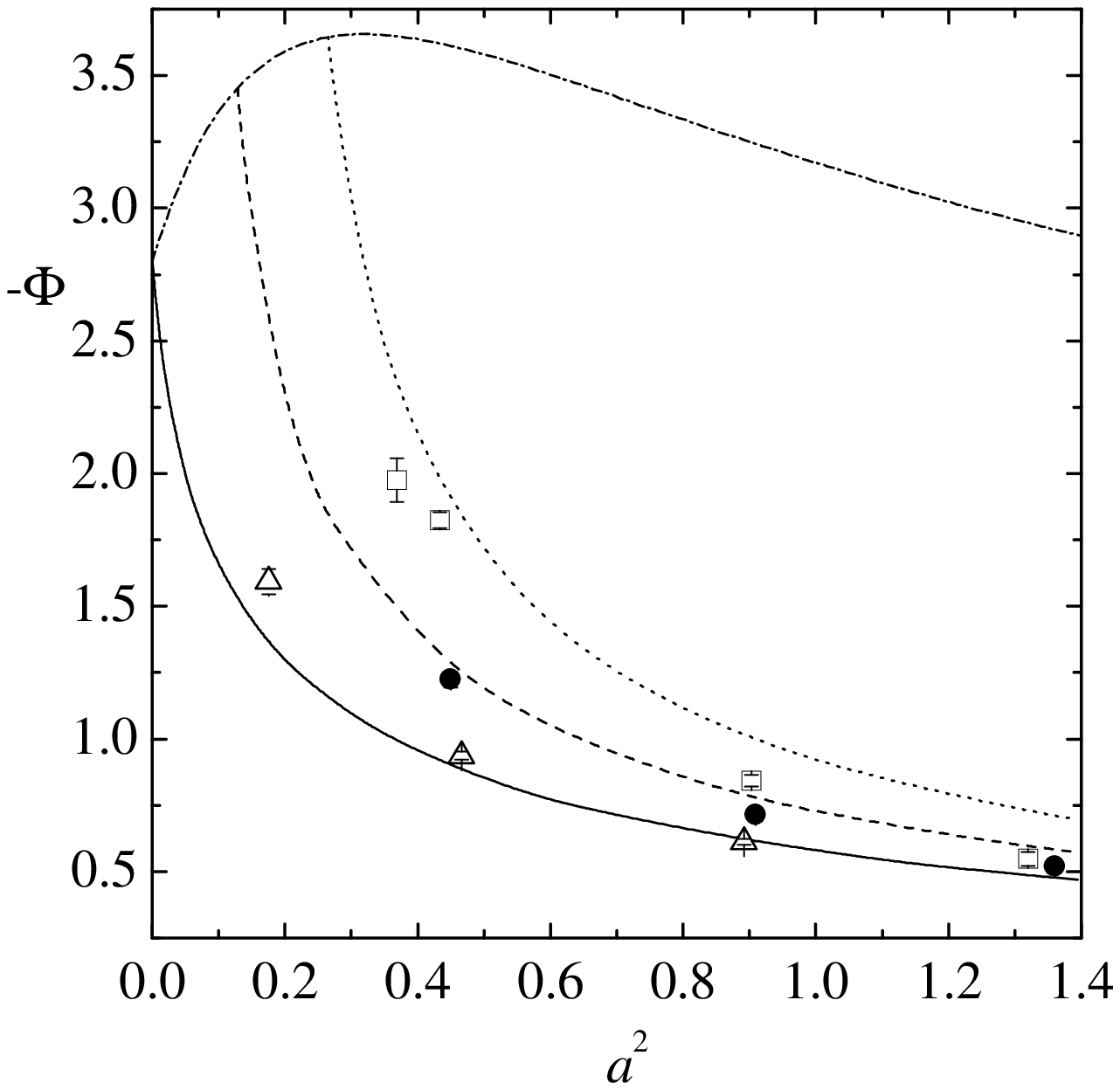}}
\caption{ The same as in Fig.\ \ref{fig4} but for the cross coefficient $
\Phi $. }
\label{fig8}
\end{center}
\end{figure}

The transport of energy is characterized by the coefficients $F_{\lambda }$
and $\Phi $, defined by Eq.\ (\ref{4.19}). They are plotted in Figs.\ \ref
{fig7} and \ref{fig8}, respectively. As happens with $F_{\eta }$, the
generalized thermal conductivity $F_{\lambda }$ decreases as the shear rate
increases. The accuracy of the kinetic model predictions for this quantity
is less satisfactory than in the case of the viscosity function. This is a
consequence of the fact that the model slightly underestimates the cooling
rate $\zeta ^{\ast }$ (Fig.\ \ref{fig1}) and the curvature parameter $\gamma 
$ (Fig.\ \ref{fig2}), while it slightly overestimates $F_{\eta }$ (Fig.\ \ref
{fig4}) and so, according to Eq.\ (\ref{4.18.3.1}), all these discrepancies
contribute to magnify the inaccuracy of the value of $F_{\lambda }$
predicted by the model as well. In the case of the coefficient $\Phi $, the
agreement between theory and simulation is quite good. This is rather
satisfactory if one takes into account that this is a coefficient measuring
complex coupling effects between the velocity and temperature gradients,
which are absent at the Navier-Stokes regime. { Also note that the
kinetic model yields $\Phi(a=0,\alpha=1)=-2.8$, while $\Phi(a=0,
\alpha=1)=-3.5$ in the Boltzmann equation. This is again due to the fact
that relaxation effects are described by a single frequency in the BGK model}.
\vspace{0.5cm}

\vspace{0.5cm}

\section{Discussion}

\label{sec6}

The exact macroscopic balance equations for mass, energy, and momentum
become closed hydrodynamic equations when the energy and momentum fluxes are
specified in terms of the hydrodynamic fields. In Section \ref{sec2.0}
specific forms for these fluxes were postulated to hold for Couette flow,
and the resulting hydrodynamic equations were solved exactly for the
temperature, pressure, and flow velocity. In Section \ref{sec2} an exact
solution to a model kinetic equation was obtained for Couette flow and the
assumed forms for the fluxes was verified. In addition, it is shown in
Appendix \ref{appC} that the hydrodynamic fields obtained directly from this
distribution function are the same as those obtained from solution to the
hydrodynamic equations. Thus, both hydrodynamic and kinetic descriptions
have been verified as exact and self-consistent. The kinetic theory provides
the distribution function in phase space, Eq.\ (\ref{c2}), as an explicit
functional of the hydrodynamic fields and therefore allows in addition the
calculation of non-hydrodynamic properties in terms of them. These fields
are given by Eqs.\ (\ref{n16}) and (\ref{n18}) with the parameter $\gamma
\left( {a},\alpha \right) $ determined by the consistency condition (\ref
{4.14}).

Some comments to summarize these results and provide perspective are as
follows:

\begin{itemize}
\item  The model kinetic equation considered is a generalization of the
familiar BGK equation used to model the Boltzmann equation for elastic
collisions, and reduces to it for $\alpha =1$. The utility of the BGK
equation to address complex states not accessible via the Boltzmann equation
is well-established. The results here and those of Refs.\ \onlinecite{BRM97}
and \onlinecite{MGSB99} confirm its value for granular media as well.
Although these kinetic model equations are structurally simple, they are
highly nonlinear due to the implicit dependence of $f_{\ell }$ on the
distribution function via the hydrodynamic fields.

\item  The kinetic model equation provides a coupled set of singular
nonlinear integral equations for the hydrodynamic fields. Incorporation of
specific boundary conditions is straightforward and an advantage of this
description. However, explicit results generally require approximations near
the homogeneous state or numerical solution. Exact analytic results valid
for large spatial gradients and arbitrary inelasticity are rare even for the
kinetic model equations. They require idealized boundary conditions such
that boundary layer complications are avoided and ``simple'' profiles are
possible in the appropriate variables.

\item  The idealized boundary conditions considered here are homogeneous,
such that the solution is ``normal'' --- all space dependence occurs
entirely through the hydrodynamic fields. Since the fluxes in macroscopic
balance equations (\ref{2.6})--(\ref{2.8}) are determined from moments of
this distribution, they can be calculated as functions of the hydrodynamic
fields. Use of these in the macroscopic balance equations then gives a
closed set of equations for the fields. The analysis here provides a non
trivial example of the existence of hydrodynamics for a strongly
inhomogeneous state together with the solution to those equations. The
solution is valid even for strong dissipation subject to the condition ${\
\gamma }\geq 0$ or ${a}\geq {a}_{c}=\sqrt{d{\zeta }^{\ast }\left( \alpha
\right) /2[1-{\zeta }^{\ast }(\alpha )]}$.

\item  {\ The results derived here include as a limiting case those
previously obtained for the uniform shear flow problem. \cite{BRM97,MGSB99}
This corresponds to an exact balance between the viscous heating and the
inelastic cooling, resulting in a uniform temperature ($\gamma\to 0^+$).
{}From that point of view, the critical shear rate $a_c$ gives the
relationship between the shear rate and the coefficient of restitution in
the uniform shear flow. }

\item  The existence of a critical shear rate ${a}_{c}$ below which the
analysis does not apply restricts the conditions such that viscous heating
dominates collisional cooling. The results do not apply, therefore, to the
simple case of zero shear rate and constant temperature walls. In this
latter case, simulations and asymptotic analysis \cite{BC98,Cordero} suggest
that the temperature profile is very close to that of (\ref{2.18}) but with $
{\ \gamma }\leq 0$. The proof for this is still not available.

\item  Although the kinetic model is only a crude representation of the
Boltzmann equation, it does preserve the most important features for
transport, such as the reference homogeneous state and the macroscopic
conservation laws. Otherwise there are only two adjustable parameters, the
collision rate $\nu $ and the cooling rate $\zeta $, to improve
correspondence with the Boltzmann equation. Here they have been chosen to
fit the Navier-Stokes order viscosity. Surprisingly, this is sufficient to
allow good agreement with the Boltzmann results for rheological properties
far from equilibrium. This is true for elastic collisions \cite{MG98,MGS00}
and now has been confirmed here for inelastic collisions.
\end{itemize}

\acknowledgments This work has been done under the auspices of the Agencia
Espa\~nola de Cooperaci\'on Internacional (Programa de Cooperaci\'on
Interuniversitaria Hispano-Marroqu\'{\i}). J. M. M., V. G., and A.S.
acknowledge partial support from the DGES (Spain) through grant No.\
PB97-1501 and from the Junta de Extremadura (Fondo Social Europeo) through
grant No.\ IPR99C031. A.S. is also grateful to the DGES (Spain) for a
sabbatical grant (No.\ PR2000-0117 0027235927). The research of J. D. W. was
supported in part by a grant from the National Science Foundation, NSF PHY
9722133.

\appendix

\section{Action of the exponential operator}

\label{appA} In this Appendix we evaluate the action of the operator $e^{- 
{\cal D}t}$, where ${\cal D}$ is given by Eq.\ (\ref{2.16}). To do this, it
is convenient to refer the velocities of the particles to a local Lagrangian
frame ${\bf V}={\bf v}-{\bf u}$ and to introduce a scaled space variable as 
\begin{equation}
s=\int_{y_{0}}^{y}dy^{\prime }\,\nu _{0}(y^{\prime }).  \label{m1}
\end{equation}
In terms of this variable, the velocity and temperature profiles become
simply 
\begin{equation}
u_{x}\left( s\right) ={a}s,\quad T\left( s\right) =T_{0}-{m{\gamma }}s^{2},
\label{2.18bis}
\end{equation}
where we have taken into account that $\lambda _{0}=(d+2)\eta _{0}/2m$ in
the BGK kinetic model. In the new Lagrangian frame the operator ${\cal D}$
becomes 
\begin{equation}
{\cal D}=V_{y}\partial _{s}-{a}V_{y}\partial _{V_{x}}-\frac{{\zeta ^{\ast }} 
}{2}{\bf V}\cdot \partial _{{\bf V}},  \label{3.1}
\end{equation}
where the derivative $\partial _{s}$ is taken at constant ${\bf V}$.

Let us consider the operators $A\equiv -V_{y}\partial _{s}$, $B\equiv 
\case{1}{2}{\zeta^* }{\bf V}\cdot \partial_{{\bf V}}$, and $C\equiv a
V_{y}\partial _{V_{x}}$. It is easy to see that the operators $A+B$ and $C$
commute, namely, 
\begin{equation}
\lbrack A+B,C]\equiv (A+B)C-C(A+B)=0 .  \label{a1}
\end{equation}
Thus, the action of the operator $e^{-{\cal D}t}\equiv e^{(A+B+C)t}$ on a
given function $F(s,{\bf V})$ is 
\begin{eqnarray}
e^{(A+B+C)t}F(s,{\bf V})&=&e^{(A+B)t}e^{Ct}F =e^{Ct}e^{(A+B)t}F  \nonumber \\
&=&e^{(A+B)t}F(s,{\bf V}+t{\sf a}\cdot {\bf V}) ,  \label{a2}
\end{eqnarray}
where ${\sf a}$ is the matrix with elements ${a}_{ij}=a\delta _{ix}\delta
_{jy}$. It remains to evaluate the action of the operator $e^{(A+B)t}$ on an
arbitrary function of the velocity, $h(s,{\bf V})$: 
\begin{equation}
e^{(A+B)t}h(s,{\bf V})\equiv e^{Bt}H(t,s,{\bf V}) .  \label{a3}
\end{equation}
An equation for $H(t,s,{\bf V})$ follows by differentiating both sides to
get 
\begin{equation}
\partial _{t}H(t,s,{\bf V})-e^{-Bt} A e^{Bt}H(t,s,{\bf V})=0.  \label{a4}
\end{equation}
Using the property for scale transformation 
\begin{equation}
e^{\frac{1}{2}{\zeta^* }t{\bf V}\cdot \partial_{{\bf V}}}\chi ({\bf V})=\chi
\left( e^{\frac{1}{2}{\zeta^* }t}{\bf V}\right) ,  \label{a5}
\end{equation}
(\ref{a4}) simplifies to 
\begin{equation}
\partial _{t}H(t,s,{\bf V})+e^{-\frac{1}{2}{\zeta^*}t}V_{y}\partial
_{s}H(t,s,{\bf V})=0.  \label{a.6}
\end{equation}
Integrating this with the initial condition $H(0,s,{\bf V})=h(s,{\bf V})$
gives 
\begin{equation}
H(t,s,{\bf V})=\exp \left[ -\frac{2}{{\zeta^* }}\left( 1-e^{-\frac{1}{ 2}{\
\zeta^* }t}\right) V_{y}\partial _{s}\right] h(s,{\bf V}).  \label{a7}
\end{equation}
Use of this in (\ref{a3}) gives the desired result 
\begin{eqnarray}
e^{(A+B)t}h(s,{\bf V})&=&\exp \left( \frac{1}{2}{\zeta^*}t{\bf V}\cdot
\partial_{{\bf V}}\right) \exp \left[ -\frac{2}{{\zeta^* }}\left( 1-e^{- 
\frac{1}{2}{\zeta^* }t}\right) V_{y}\partial _{s}\right] h(s,{\bf V}) 
\nonumber \\
&=&\exp \left[ -\tau (t)V_{y}\partial _{s}\right] h(s,e^{\frac{1}{2} {\
\zeta^* }t}{\bf V})=h\left(s-\tau (t)V_{y},e^{\frac{1}{2}{\zeta^* }t}{\bf V}
\right),  \label{a8a}
\end{eqnarray}
where 
\begin{equation}
\tau (t)=\frac{2}{{\zeta^* }}\left( \ e^{\frac{1}{2}{\zeta^* } t}-1\right) .
\label{3.3}
\end{equation}
Finally, combining Eqs.\ (\ref{a2}) and (\ref{a8a}) gives the result 
\begin{equation}
e^{-{\cal D}t}F(s,{\bf V})= e^{(A+B+C)t}F(s,{\bf V})=F\left(s-\tau
(t)V_{y},e^{\frac{1}{2}{\zeta^* } t}\left( {\bf V}+t{\sf a}\cdot{\bf V}
\right) \right).  \label{a9}
\end{equation}

\section{Idealized boundary conditions}

\label{appB}

The boundary conditions for the general solution (\ref{2.15}) are imposed by
the choice of $f_{B}$, which is a solution to 
\begin{equation}
\left( 1-\zeta ^{\ast }-\frac{d}{2}{\zeta ^{\ast }}+{\cal D}\right) f_{B}=0
\label{b1}
\end{equation}
or, equivalently, 
\begin{equation}
\partial _{s}f_{B}=-V_{y}^{-1}\left( 1-\zeta ^{\ast }-\frac{d}{2}{\zeta
^{\ast }}-{a}V_{y}\partial _{V_{x}}-\frac{{\zeta ^{\ast }}}{2}{\bf V}\cdot
\partial _{{\bf V}}\right) f_{B},
\end{equation}
where we have made use of Eq.\ (\ref{3.1}). This can be integrated to get 
\begin{eqnarray}
f_{B}(s,{\bf V)} &=&\Theta (V_{y})\exp \left[ -\frac{s+{\cal L}_{-}}{V_{y}}
\left( 1-\zeta ^{\ast }-\frac{d}{2}{\zeta ^{\ast }}-{a}V_{y}\partial
_{V_{x}}-\frac{{\zeta ^{\ast }}}{2}{\bf V}\cdot \partial _{{\bf V}}\right) 
\right] \Phi _{-}({\bf V})  \nonumber \\
&&+\Theta (-V_{y})\exp \left[ -\frac{s-{\cal L}_{+}}{V_{y}}\left( 1-\zeta
^{\ast }-\frac{d}{2}{\zeta ^{\ast }}-{a}V_{y}\partial _{V_{x}}-\frac{{\zeta
^{\ast }}}{2}{\bf V}\cdot \partial _{{\bf V}}\right) \right] \Phi _{+}({\bf 
V }).  \label{b3}
\end{eqnarray}
Here $\Theta (x)$ is the Heaviside step function, $s=\pm {\cal L}_{\pm }$
correspond to the location of the walls at $y=\pm L$, and $\Phi _{\pm }({\bf 
V})$ are arbitrary functions. These functions are fixed by the boundary
conditions at $s=\pm {\cal L}_{\pm }$: 
\begin{equation}
\Theta (\pm V_{y})f(\mp {\cal L}_{\mp },{\bf V)}=\Theta (\pm V_{y})\Phi
_{\mp }({\bf V}).  \label{b4}
\end{equation}

Idealized boundary conditions are constructed as follows. First we write the
boundary conditions as 
\begin{equation}
\Theta (\pm V_{y})f(\mp{\cal L}_{\mp},{\bf V})=\Theta (\pm V_{y}) \frac{
n_{\mp}}{C_{\mp}}(m/2 T_{\mp})^{d/2}\varphi_{\mp}\left({\bf V}/\sqrt{2
T_\mp/m}\right),  \label{b8}
\end{equation}
where 
\begin{equation}  \label{b5}
n_{\mp}\equiv \left(2 T_{\mp}/m\right)^{-1/2} \int d{\bf V}\, \Theta(\mp
V_y)|V_y| f(\mp{\cal L}_{\mp},{\bf V}),
\end{equation}
\begin{equation}  \label{b6}
C_{\mp}\equiv \int d\bbox{\xi}\, \Theta(\pm \xi_y)|\xi_y| \varphi_{\mp}( 
\bbox{\xi}),
\end{equation}
$\varphi_{\mp}(\bbox{\xi})$ being the distributions of velocities away from
the walls normalized as 
\begin{equation}  \label{b7}
\int d\bbox{\xi}\, \varphi_{\mp}(\bbox{\xi})=\frac{2}{d} \int d\bbox{\xi}\,
\xi^2\varphi_{\mp}(\bbox{\xi})=1.
\end{equation}
The usual choice is the Maxwell-Boltzmann distribution $\varphi_+(\bbox{\xi}
) =\varphi_-(\bbox{\xi})=\pi^{-d/2}\exp(-\xi^2)$. Now consider the limit of
infinitely cold walls, $T_{\pm}\to 0$. In that limit $\Phi_\pm({\bf V})\to 0$
since $\varphi_{\pm}(\bbox{\xi})$ must vanish if $|\bbox{\xi}|\to \infty$.
Consequently, 
\begin{equation}
\lim_{T_{\pm}\to 0}f_{B}(s,{\bf V)} =0.  \label{b9}
\end{equation}

\section{Consistency conditions and calculation of the fluxes}

\label{appC}

\subsection{Generating function}

\label{appC1} The consistency conditions for the assumed forms of the
hydrodynamic fields are the $d+2$ equations of (\ref{2.3}). Evaluation of
the right hand sides of these equations requires the explicit form for the
distribution function in terms of these fields. Using the results of
Appendix \ref{appA}, the solution (\ref{2.15}) with the idealized boundary
conditions discussed above becomes 
\begin{eqnarray}
f(s,{\bf V})&=& (1-\zeta^*)\int_{0}^{\infty }dt\ e^{-(1-\zeta^*-\frac{d}{2}{
\zeta^* })t}\ \ e^{{a}tV_{y}\partial _{V_{x}}}e^{-\tau(t) V_{y}\partial
_{s}}\ f_{\ell }(s,\ e^{\frac{1}{2}{\zeta^* }t}{\bf V})  \nonumber \\
&=& (1-\zeta^*)\int_{0}^{\infty }dt\ e^{-(1-\zeta^*-\frac{d}{2}{\ \zeta^* }
)t} f_{\ell }\left(s-V_{y}\tau (t),\ e^{\frac{1}{2}{\zeta^* } t}\left( {\bf V
}+ t{\sf a}\cdot{\bf V}\right) \right) ,  \label{c2}
\end{eqnarray}
where the local equilibrium distribution is 
\begin{equation}
f_{\ell }(s,{\bf V})=\frac{2p}{m}\Theta
(T(s))v_{0}^{-(d+2)}(s)\pi^{-d/2}\exp [-\left( V/v_{0}(s)\right) ^{2}]
\label{c3}
\end{equation}
with $v_{0}^{2}(s)=2 T(s)/m$.

The velocity integrals for the consistency conditions and for the fluxes are
all low order moments of this distribution function. They can be obtained
from appropriate derivatives of the generating function 
\begin{equation}
G(s,{\bf k})=\int d{\bf V}e^{i{\bf k\cdot V}}f(s,{\bf V}).  \label{3.4}
\end{equation}
Making use of the solution (\ref{c2}), the generating function becomes 
\begin{eqnarray}  \label{c4}
G(s,{\bf k})&=&(1-\zeta^*)\int_{0}^{\infty }dt\, e^{-(1-\zeta^*-\frac{d}{2}{
\ \zeta^* } )t}\int d{\bf V}e^{i{\bf k\cdot V}}e^{{a}tV_{y}\partial
_{V_{x}}}e^{-\tau(t) V_{y}\partial _{s}}\ f_{\ell }(s, e^{\frac{1}{2} {\
\zeta^* }t}{\bf V})  \nonumber \\
&=&(1-\zeta^*)\int_{0}^{\infty }dt\ e^{-(1-\zeta^*)t}\int d{\bf V}e^{i{\bf k}
^{\prime}(t){\bf \cdot V} }e^{-\tau_{1}(t)V_{y}\partial _{s}} f_{\ell }(s, 
{\bf V}),
\end{eqnarray}
where 
\begin{equation}
{\bf k}^{\prime}(t)=e^{-\frac{1}{2}{\zeta^*}t} e^{-{a}tk_x\partial_{k_y}} 
{\bf k},\quad \tau _{1}(t)=\frac{2}{{\zeta^* }}\left( 1-e^{-\frac{1}{2}{\
\zeta^* }t}\right).  \label{c5}
\end{equation}
In the last step of Eq.\ (\ref{c4}) we have made the change ${\bf V}\to e^{
\frac{1}{2}{\zeta^*}t}{\bf V}$ and have used the general property 
\begin{equation}  \label{c4bis}
\int d{\bf V}\, F_1({\bf V})e^{{a}tV_{y}\partial_{V_{x}}}F_2({\bf V})= \int
d {\bf V}\, F_2({\bf V})e^{-{a}tV_{y}\partial_{V_{x}}}F_1({\bf V}).
\end{equation}
Integration over ${\bf V}_\perp\equiv {\bf V}-V_y \widehat{{\bf y}}$ yields 
\begin{equation}  \label{c6}
G(s,{\bf k})=G_+(s,{\bf k})+G_-(s,{\bf k}),
\end{equation}
where 
\begin{equation}  \label{c7}
G_{\pm}(s,{\bf k})= \frac{2p}{m}(1-\zeta^*)\int_{0}^{\infty
}dV_{y}\int_{0}^{\infty }dt\, e^{-(1-\zeta^*)t}e^{\pm
ik_{y}^{\prime}(t)V_{y}}e^{\mp
\tau_1(t)V_y\partial_s}e^{-V_y^2/v_0^2(s)}F(s, {\bf k}_\perp^{\prime}(t)),
\end{equation}
with 
\begin{equation}
F(s,{\bf k}_\perp^{\prime}(t))= \Theta (v_{0}^{2}(s))v_{0}^{-3}(s)\pi
^{-1/2}\exp \left[-\frac{1}{4}{k_\perp^{\prime}}^2(t) v_{0}^{2}(s)\right].
\label{c8}
\end{equation}
We will focus on $G_{+}(s,{\bf k})$ since $G_{-}(s,{\bf k})$ can be
evaluated as $G_{-}(s,{\bf k})=G_{+}(-s,-{\bf k})$.

Next, define the change of variables in the $t$ integration 
\begin{equation}  \label{c9}
z=V_{y}\tau _{1}(t) ,\quad dz=\left(V_{y}-\frac{{\ \zeta^* }z}{2}\right)
dt,\quad t=-\frac{2}{{\zeta^* }}\ln \left( 1-\frac{{\zeta^* }z}{2V_{y}}
\right) .
\end{equation}
The function $G_+(s,{\bf k})$ becomes 
\begin{eqnarray}  \label{c10}
G_+(s,{\bf k})&=&\frac{2p}{m}(1-\zeta^*)\int_{0}^{\infty
}dV_{y}\int_{0}^{\infty}dz\,
e^{-(1-\zeta^*)t}e^{ik_{y}^{\prime}(t)V_{y}}\left( V_{y}-\frac{{\zeta^* }}{2}
z\right) ^{-1}\Theta \left(V_{y}-\frac{{\zeta^* }}{2}z\right)  \nonumber \\
&&\times e^{-V_y^2/v_0^2(s-z)} F(s-z,{\bf k}_\perp^{\prime}(t))  \nonumber \\
&=&\frac{2p}{m}(1-\zeta^*)\int_{0}^{\infty }du\,
e^{-u^{2}}u^{-1}\int_0^\infty dz\, e^{-(1-\zeta^*)t}e^{ik_{y}^{\prime}(t)u
v_{0}(s-z)}\left[ 1-\frac{{\zeta^* }}{2}\frac{z}{uv_{0}(s-z)}\right] ^{-1} 
\nonumber \\
&&\times \Theta \left( 1-\frac{{\zeta^* }}{2}\frac{z}{uv_{0}(s-z)}\right)
F(s-z,{\bf k}_\perp^{\prime}(t)).
\end{eqnarray}
A change of variables in the $V_{y}$ integral has been made, $
V_{y}\rightarrow u v_{0}(s-z)$. Accordingly, the variable $t$ becomes 
\begin{equation}  \label{c11}
t=-\frac{2}{{\zeta^* }}\ln \left[ 1-\frac{{\zeta^* }z}{2uv_{0}(s-z)}\right].
\end{equation}

Next, for the $z$ integral change variables to 
\begin{equation}  \label{c12}
w=\frac{z}{v_{0}(s-z)}.
\end{equation}
As a consequence, $z$ as a function of $w$ is $z=z_+(s,w)$, where $
z_{+}(s,w) $ is the positive root of the quadratic equation obtained by
squaring (\ref{c12}): 
\begin{equation}
\left( 1+2{\gamma }w^{2}\right) z^{2}-4{\gamma } w^{2}sz-w^{2}v_{0}^{2}(s)=0.
\label{c13}
\end{equation}
It can be easily shown that 
\begin{equation}
\frac{dz}{dw}=v_0^{-2}(s) \left(\frac{z_+}{w}\right) ^{3}\left[ 1-\frac{
z_{+} }{2}\partial _{s}\ln T(s) \right] ^{-1}.  \label{c14}
\end{equation}
The generating function is now 
\begin{eqnarray}  \label{c15}
G_+(s,{\bf k})&=&\frac{2p}{mv_{0}^{2}(s)\pi ^{1/2}}(1-\zeta^*)\int_{0}^{
\infty }du\,e^{-u^{2}}u^{-1}\int_{0}^{\infty }dw\,e^{-(1-\zeta^*)t}\left( 1- 
\frac{{\zeta^* } }{2}\frac{w}{u}\right)^{-1}\Theta \left( 1-\frac{{\zeta^* } 
}{2}\frac{ w}{u}\right)  \nonumber \\
&&\times e^{ik_{y}^{\prime}(t)z_{+}u/w}\left[ 1-\frac{z_{+}}{2}\partial
_{s}\ln T(s)\right] ^{-1} e^{-{k_\perp^{\prime}}^2(t) (z_{+} / 2w)^2},
\end{eqnarray}
where $t$ is now a function of $u$ and $w$, 
\begin{equation}
t=-\frac{2}{{\zeta^* }}\ln \left( 1-\frac{{\zeta^* } w}{2u}\right).
\label{c16}
\end{equation}
Finally, change variables in the $w$ integration to $t$, so that 
\begin{equation}  \label{c17}
w=\frac{2u}{{\zeta^* }}\left( 1-e^{-\frac{1}{2}{\zeta^* } t}\right) ,\quad
dt=u^{-1}\left( 1-\frac{{\zeta^* }w}{2u} \right) ^{-1}dw.
\end{equation}
This leads to the explicit result 
\begin{equation}
G_+(s,{\bf k})=\frac{p}{ T(s)\pi ^{1/2}}(1-\zeta^*)\int_{0}^{\infty
}du\,e^{-u^{2}}\int_{0}^{\infty }dt\,e^{-(1-\zeta^*)t}A_+(s,{\bf k},u,t),
\label{c20}
\end{equation}
where 
\begin{equation}  \label{c21}
A_+(s,{\bf k},u,t)=\left[ 1-\frac{z_+(s,w)}{2}\partial _{s}\ln T(s)\right]
^{-1}\exp\left[i\left( k_{y}- {a}tk_{x}\right)\frac{z_+(s,w)}{\tau(t)} - 
\frac{k_\perp^2}{4}\frac{z_+^2(s,w)}{u^2\tau^2(t)}\right].
\end{equation}
It must be recalled that $w$ is a function of $u$ and $t$ given by (\ref{c17}).

The function $G_-(s,{\bf k})=G_+(-s,-{\bf k})$ is then 
\begin{equation}  \label{c22}
G_-(s,{\bf k})=\frac{p}{ T(s)\pi ^{1/2}}(1-\zeta^*)\int_{0}^{\infty
}du\,e^{-u^{2}}\int_{0}^{\infty }dt\,e^{-(1-\zeta^*)t}A_-(s,{\bf k},u,t),
\end{equation}
with 
\begin{equation}
A_-(s,{\bf k},u,t)=\left[ 1-\frac{z_-(s,w)}{2}\partial _{s}\ln T(s)\right]
^{-1}\exp\left[i\left( k_{y}- {a}tk_{x}\right)\frac{z_-(s,w)}{\tau(t)} - 
\frac{k_\perp^2}{4}\frac{z_-^2(s,w)}{u^2\tau^2(t)}\right],
\end{equation}
where $z_-(s,w)=-z_+(-s,w)$ is the negative root of the quadratic equation 
(\ref{c13}). Real roots require 
\begin{equation}
\left( 4{\gamma }sw^{2}\right) ^{2}+4\left( 1+2{\gamma } w^{2}\right) 2w^{2} 
\frac{ T(s)}{m}\geq 0,  \label{3.8}
\end{equation}
which is satisfied for positive $T(s)$ {\em and\/} positive ${\gamma}$.

\subsection{Consistency conditions}

\label{appC2}

The consistency condition for the density becomes 
\begin{eqnarray}
n(s) &=&\int d{\bf v}f=G(s,{\bf k=0})  \nonumber \\
&=&\frac{p}{T(s)\pi ^{1/2}}(1-\zeta ^{\ast })\int_{0}^{\infty
}du\,e^{-u^{2}}\int_{0}^{\infty }dt\,e^{-(1-\zeta ^{\ast })t}\left\{ \left[
1-\frac{z_{+}}{2}\partial _{s}\ln T(s)\right] ^{-1}+\left[ 1-\frac{z_{-}}{2}
\partial _{s}\ln T(s)\right] ^{-1}\right\}  \nonumber \\
&=&\frac{p}{T(s)}.  \label{3.9}
\end{eqnarray}
Use has been made of the identity 
\begin{equation}
\left[ 1-\frac{z_{+}}{2}\partial _{s}\ln T(s)\right] ^{-1}+\left[ 1-\frac{
z_{-}}{2}\partial _{s}\ln T(s)\right] ^{-1}=2,  \label{3.10}
\end{equation}
which follows from the explicit forms for the roots of (\ref{c13}). The
result (\ref{3.9}) is consistent with the required equation of state
defining $p$.

The consistency condition for the temperature is 
\begin{equation}
dn(s)T(s)=m\int d{\bf v}V^{2}f=-m\left[ \partial _{{\bf k}}^{2}G(s,{\bf k}) 
\right] _{{\bf k}={\bf 0}}.  \label{3.11}
\end{equation}
Direct evaluation gives 
\begin{eqnarray}
1 &=&\frac{m}{dT(s)\pi ^{1/2}}(1-\zeta ^{\ast })\int_{0}^{\infty
}du\,e^{-u^{2}}\int_{0}^{\infty }dt\,e^{-t}\left[ \frac{d-1}{2}+\left(
1+a^{2}t^{2}\right) u^{2}\right] w^{-2}  \nonumber \\
&&\times \left\{ z_{+}^{2}\left[ 1-\frac{z_{+}}{2}\partial _{s}\ln T(s) 
\right] ^{-1}+z_{-}^{2}\left[ 1-\frac{z_{-}}{2}\partial _{s}\ln T(s)\right]
^{-1}\right\} .
\end{eqnarray}
Using the identity 
\begin{equation}
z_{+}^{2}\left[ 1-\frac{z_{+}}{2}\partial _{s}\ln T(s)\right]
^{-1}+z_{-}^{2} \left[ 1-\frac{z_{-}}{2}\partial _{s}\ln T(s)\right] ^{-1}=
\frac{4T(s)}{m} \frac{w^{2}}{1+2{\gamma }w^{2}},  \label{3.13}
\end{equation}
one gets the final result 
\begin{equation}
1=\frac{4}{d\pi ^{1/2}}(1-\zeta ^{\ast })\int_{0}^{\infty
}du\,e^{-u^{2}}\int_{0}^{\infty }dt\,e^{-t}\left[ \frac{d-1}{2}+\left( 1+{a}
^{2}t^{2}\right) u^{2}\right] \left( 1+2{\gamma }w^{2}\right) ^{-1}.
\label{3.16}
\end{equation}
This condition is enforced by using it as the definition of ${\gamma }$.

Finally, the consistency conditions for the flow velocity components are 
\begin{equation}
{\ 0}=m\int d{\bf v}V_{i}f=\left[ \partial _{k_{i}}G(s,{\bf k})\right] _{
{\bf k}={\bf 0}}.  \label{3.18}
\end{equation}
This is clearly satisfied for the $z$ component. The $x$ and $y$ components
are proportional to integrals including the term 
\begin{equation}
z_{+}\left[ 1-\frac{z_{+}}{2}\partial _{s}\ln T(s)\right] ^{-1}+z_{-}\left[
1-\frac{z_{-}}{2}\partial _{s}\ln T(s)\right] ^{-1},  \label{3.19}
\end{equation}
which identically vanishes.

This completes confirmation of the consistency conditions for the
hydrodynamic fields. In summary, the distribution function (\ref{c2}) with
the hydrodynamic fields (\ref{2.18}) constitutes an exact solution to the
kinetic equation.

\subsection{Calculation of the fluxes}

\label{appC3}

The macroscopic transport properties of the steady Couette flow are given by
the momentum and heat fluxes, Eq.\ (\ref{2.9}). In terms of the generating
function these are 
\begin{equation}
P_{ij}=-m\left[ \partial _{k_{i}}\partial _{k_{j}}G(s,{\bf k})\right] _{{\bf 
k}={\bf 0}},\quad {\bf q}(s)=i\frac{1}{2}m\left[ \partial_{{\bf k}}
\partial_{{\bf k}}^2 G(s,{\bf k})\right] _{{\bf k}={\bf 0}}.  \label{4.1}
\end{equation}
Following similar steps as in the consistency condition for the temperature,
the nonzero elements of the pressure tensor can be evaluated. They are given
by 
\begin{equation}
P_{xx}=\frac{2p}{\pi ^{1/2}}(1-\zeta^*)\int_{0}^{\infty
}du\,e^{-u^{2}}\int_{0}^{\infty }dt\,e^{- t}\left[ 1+2\left( {a} tu\right)
^{2}\right] \left( 1+2 {\gamma }w^{2}\right) ^{-1},  \label{4.2}
\end{equation}
\begin{equation}
P_{yy}=\frac{4p}{\pi ^{1/2}}(1-\zeta^*)\int_{0}^{\infty }du\,e^{-u^{2}}u^{2}
\int_{0}^{\infty }dt\,e^{- t}\left( 1+2 {\gamma } w^{2}\right) ^{-1},
\label{4.3}
\end{equation}
\begin{equation}
P_{zz}=\frac{2p}{\pi ^{1/2}}(1-\zeta^*)\int_{0}^{\infty
}du\,e^{-u^{2}}\int_{0}^{\infty }dt\,e^{-t}\left( 1+2 {\gamma } w^{2}\right)
^{-1},  \label{4.4}
\end{equation}
\begin{equation}
P_{xy}=-\frac{4 {a}p}{\pi ^{1/2}}(1-\zeta^*)\int_{0}^{\infty
}du\,e^{-u^{2}}u^{2}\int_{0}^{\infty }dt\,e^{-t}t\left( 1+2 {\gamma }
w^{2}\right) ^{-1}.  \label{4.5}
\end{equation}
It is readily verified that these results satisfy the relationship $p=\frac{
1 }{d}\left[ P_{xx}+P_{yy}+(d-2)P_{zz}\right]$.

Let us consider now the heat flux vector. Note that, although the
temperature gradient is only directed along the $y$ direction, the presence
of the shear flow induces a nonzero $x$ component of the heat flux. Thus the
nonzero components are $q_{x}$ and $q_{y}$. Using (\ref{4.1}), $q_y$ is
found to be 
\begin{eqnarray}  \label{4.3.1}
q_{y}(s)&=&\frac{mp}{2 T(s)\pi ^{1/2}}(1-\zeta^*)\int_{0}^{\infty
}du\,e^{-u^{2}}u\int_{0}^{\infty }dt\,e^{-\left( 1+\frac{1}{2}{\zeta^* }
\right) t}w^{-3}\left[ \frac{d-1}{2}+\left( 1+ {a} ^{2}t^{2}\right)u^{2} 
\right]  \nonumber \\
&&\times \left\{ z_{+}^{3}\left[ 1-\frac{z_{+}}{2}\partial _{s}\ln T(s) 
\right] ^{-1}+z_{-}^{3}\left[ 1-\frac{z_{-}}{2}\partial _{s}\ln T(s)\right]
^{-1}\right\}  \nonumber \\
&=&\frac{8 {\gamma }ps}{\pi ^{1/2}}(1-\zeta^*)\int_{0}^{\infty
}du\,e^{-u^{2}}u\int_{0}^{\infty }dt\,e^{-\left( 1+\frac{1}{2}{\zeta^* }
\right) t}\left[ \frac{d-1}{2}+\left( 1+ {a}^{2}t^{2}\right) u^{2} \right]
w\left( 1+2 {\gamma }w^{2}\right) ^{-2} ,
\end{eqnarray}
where use has been made of the identity 
\begin{equation}  \label{4.3.2}
z_{+}^{3}\left[ 1-\frac{z_{+}}{2}\partial _{s}\ln T(s)\right]
^{-1}+z_{-}^{3} \left[ 1-\frac{z_{-}}{2}\partial _{s}\ln T(s)\right]
^{-1}=16 {\gamma}s \frac{ T(s)}{m}\frac{w^4}{(1+2 {\gamma}w^2)^2}.
\end{equation}
This can be written as 
\begin{eqnarray}
q_y&=&-\frac{2ps}{\pi ^{1/2}}(1-\zeta^*)\int_{0}^{\infty
}du\,e^{-u^{2}}\int_{0}^{\infty }dt\,e^{-t}\left[ \frac{d-1}{2}+\left( 1+ {a}
^{2}t^{2}\right)u^{2} \right] \frac{\partial}{\partial {t}} \left( 1+2 {\
\gamma }w^{2}\right) ^{-1}  \nonumber \\
&=&-\frac{dp{\zeta^* }}{2}s+\frac{4 {a}^{2}ps}{\pi ^{1/2}}(1-\zeta^*)
\int_{0}^{\infty }du\,e^{-u^{2}}u^{2}\int_{0}^{\infty }dt\,e^{-t}t\left( 1+2 
{\gamma }w^{2}\right) ^{-1}.  \label{4.9}
\end{eqnarray}
An integration by parts in the $t$ integral and the consistency condition 
(\ref{3.16}) have been used to obtain the last equality. Comparison of this
result with that for $P_{xy}$ above shows 
\begin{equation}
q_{y}=\frac{1}{2m {\gamma}}\left( \frac{dp{\zeta^* }}{2}+ {a}P_{xy}\right)
\partial_s T.  \label{4.10}
\end{equation}
This result is in fact required by the hydrodynamic equations to support the
forms for the hydrodynamic fields.

In a similar way $q_{x}(s)$ is calculated, 
\begin{eqnarray}
q_{x}(s)&=&-\frac{{a}mp}{2 T(s)\pi ^{1/2}}(1-\zeta^*)\int_{0}^{\infty
}du\,e^{-u^{2}}u\int_{0}^{\infty }dt\,e^{-\left( 1+\frac{1}{2}{\zeta^* }
\right) t}tw^{-3}\left[ \frac{d+1}{2}+\left( 1+ {a}^{2}t^{2}\right) u^{2} 
\right]  \nonumber \\
&&\times \left\{ z_{+}^{3}\left[ 1-\frac{z_{+}}{2}\partial _{s}\ln T(s) 
\right] ^{-1}+z_{-}^{3}\left[ 1-\frac{z_{-}}{2}\partial _{s}\ln T(s)\right]
^{-1}\right\}  \nonumber \\
&=& \frac{4 p}{m\pi^{1/2}} {a}(1-\zeta^*)\int_{0}^{\infty
}du\,e^{-u^{2}}u\int_{0}^{\infty }dt\,e^{-\left( 1+\frac{1}{2}{\zeta^* }
\right) t}tw\left[ \frac{d+1}{2}+\left( 1+ {a}^{2}t^{2}\right) u^{2} \right]
\nonumber \\
&&\times \left( 1+2 {\gamma }w^{2}\right) ^{-2}\partial_s T.  \label{4.11}
\end{eqnarray}

\section{Transport coefficients in the uniform shear flow limit}

\label{appD}

In this Appendix we derive the explicit expressions for the transport
coefficients along the critical shear rate $a_{c}(\alpha )$. They are
obtained by taking the limit $\gamma \rightarrow 0^{+}$ in the corresponding
expressions of Sec.\ \ref{sec2}. Equations (\ref{4.15a})--(\ref{4.18.4})
simply yield 
\begin{equation}
F_{\eta }(a_{c},\alpha )=1-\zeta ^{\ast }=\frac{d}{d+2a_{c}^{2}},  \label{D1}
\end{equation}
\begin{equation}
\Psi _{1}(a_{c},\alpha )=-2(1-\zeta ^{\ast })=-\frac{2d}{d+2a_{c}^{2}},
\label{D2}
\end{equation}
\begin{equation}
\Psi _{2}(a_{c},\alpha )=0,  \label{D3}
\end{equation}
\begin{equation}
\Phi (a_{c},\alpha )=-\frac{2}{d+2}\frac{(1-\zeta ^{\ast })(4+3\zeta ^{\ast
})}{(1+\zeta ^{\ast })^{4}(2+\zeta ^{\ast })^{4}}\left[ 7(2+3\zeta ^{\ast }+{
\ \zeta ^{\ast }}^{2})^{2}+18a_{c}^{2}(8+12\zeta ^{\ast }+5{\zeta ^{\ast }}
^{2})\right] .  \label{D4}
\end{equation}
In order to obtain $F_{\lambda }$ from Eq.\ (\ref{4.18.3.1}), we need to
evaluate $a^{2}$ and $F_{\eta }$ to first order in $\gamma $. {}From Eqs.\ 
(\ref{4.14}) and (\ref{4.15a}) one gets 
\begin{equation}
a^{2}=\frac{d}{2}\frac{\zeta ^{\ast }}{1-\zeta ^{\ast }}+A(\zeta ^{\ast
})\gamma +\cdots ,  \label{D5}
\end{equation}
\begin{equation}
F_{\eta }=1-\zeta ^{\ast }-B(\zeta ^{\ast })\gamma +\cdots ,  \label{D6}
\end{equation}
where 
\begin{eqnarray}
A(\zeta ^{\ast }) &=&\frac{2}{(1+\zeta ^{\ast })^{3}(2+\zeta ^{\ast })^{3}} 
\left[ (d+2)(2+3\zeta ^{\ast }+{\zeta ^{\ast }}^{2})^{2}\right.  \nonumber \\
&&\left. +6a_{c}^{2}(24+48\zeta ^{\ast }+33{\zeta ^{\ast }}^{2}+9{\zeta
^{\ast }}^{3}+{\zeta ^{\ast }}^{4})\right] ,  \label{D7}
\end{eqnarray}
\begin{equation}
B(\zeta ^{\ast })=12\frac{1-\zeta ^{\ast }}{(1+\zeta ^{\ast })^{2}(2+\zeta
^{\ast })^{2}}(6+6{\zeta ^{\ast }}+{\zeta ^{\ast }}^{2}).  \label{D8}
\end{equation}
Thus, the thermal conductivity is 
\begin{equation}
F_{\lambda }(a_{c},\alpha )=\frac{(1-\zeta ^{\ast })A(\zeta ^{\ast
})-a_{c}^{2}B(\zeta ^{\ast })}{d+2}.  \label{D9}
\end{equation}

Equations (\ref{D1})--(\ref{D3}) coincide with those previously derived in
Ref.\ \onlinecite{BRM97}. However, Eqs.\ (\ref{D4}) and (\ref{D9}) are new
results. Despite the fact that there is no heat flux in the uniform shear
flow, Eqs.\ (\ref{D4}) and (\ref{D9}) are intrinsic transport coefficients
characterizing the state of the system.


\begin{references}
\bibitem{C90}  C. S. Campbell, {\em Ann. Rev. Fluid Mech.}:{\bf 22}, 57
(1990).

\bibitem{Simulation}  J. J. Brey, M. J. Ruiz-Montero, and D. Cubero, {\em 
Phys. Rev. E}:{\bf 54}, 3664 (1996); J. J. Brey, D. Cubero, and M. J.
Ruiz-Montero, {\em ibid.}:{\bf 59}, 1256 (1999); S. Luding, M. M\"uller, and
S. McNamara, in {\em World Congress on Particle Technology} (Brighton, 1998,
CD: ISBN 0-85295-401-9); J. M. Montanero and A. Santos, {\em Gran. Matt.}:{\bf 
2}, 53 (2000) and cond-mat/0002323.

\bibitem{BDKS98}  J. J. Brey, J. W. Dufty, C. S. Kim, and A. Santos, {\em 
Phys. Rev. E}:{\bf 58}, 4638 (1998).

\bibitem{SG98}  N. Sela and I. Goldhirsch, {\em J. Fluid Mech.}:{\bf 361},
41 (1998).

\bibitem{GD99}  V. Garz\'o and J. W. Dufty, {\em Phys. Rev. E}:{\bf 59},
5895 (1999).

\bibitem{BDS99}  J. J. Brey, J. W. Dufty, and A. Santos, {\em J. Stat. 
Phys.}:{\bf 97}, 281 (1999).

\bibitem{Z79}  R. Zwanzig, {\em J. Chem. Phys.}:{\bf 71}, 4416 (1979).

\bibitem{SBG86}  A. Santos, J J. Brey, and V. Garz\'o, {\em Phys. Rev. A}:{\bf 
34}, 5047 (1986).

\bibitem{BSD87}  J. J. Brey, A. Santos, and J. W. Dufty, {\em Phys. Rev. A}:{\bf 
36}, 2842 (1987).

\bibitem{KDSB89}  C. S. Kim, J. W. Dufty, A. Santos, and J. J. Brey, {\em 
Phys. Rev. A}:{\bf 40}, 7165 (1989).

\bibitem{MG98}  J. M. Montanero and V. Garz\'o, {\em Phys. Rev. E}:{\bf 58},
1836 (1998).

\bibitem{BRM97}  J. J. Brey, M. J. Ruiz-Montero, and F. Moreno, {\em Phys.
Rev. E}:{\bf 55}, 2846 (1997).

\bibitem{MGSB99}  J. M. Montanero, V. Garz\'o, A. Santos, and J. J. Brey, 
{\em J. Fluid Mech.}:{\bf 389}, 391 (1999).

\bibitem{JS83}  J. T. Jenkins and S. B. Savage, {\em J. Fluid Mech.}:{\bf 
130 }, 187 (1983); J. T. Jenkins and M. W. Richman, {\em Phys. Fluids}:{\bf 
28}, 3485 (1985); {\em J. Fluid Mech.}:{\bf 192}, 313 (1988).

\bibitem{LSJC84}  C. K. K. Lun, S. B. Savage, D. J. Jeffrey, and N.
Chepurniy, {\em J. Fluid Mech.}:{\bf 140}, 223 (1984).

\bibitem{HS92}  M. A. Hopkins and H. H. Shen, {\em J. Fluid Mech.}:{\bf 244}, 
477 (1992).

\bibitem{SGN96}  I. Goldhirsch and M. L. Tan, {\em Phys. Fluids}:{\bf 8},
1753 (1996); N. Sela, I. Goldhirsch, and S. H. Noskowicz, {\em ibid.}:{\bf 8}, 
2337 (1996).

\bibitem{C00}  C. Cercignani, ``Shear flow of a granular material,'' {\em J.
Stat. Phys.}, to be published.

\bibitem{RC88}  M. W. Richman and C. S. Chou, {\em J. Appl. Math. Phys.}:{\bf 
39}, 885 (1988).

\bibitem{HJR88}  T. N. Hanes, J. T. Jenkins, and M. W. Richman, {\em J.
Appl. Mech.}:{\bf 55}, 969 (1988).

\bibitem{L96}  C. K. K. Lun, {\em Phys. Fluids}:{\bf 8}, 2868 (1996).

\bibitem{B97}  M. Babic, {\em Phys. Fluids}:{\bf 9}, 2486 (1997).

\bibitem{CC70}  S. Chapman and T. G. Cowling, {\em The Mathematical Theory
of Nonuniform Gases} (Cambridge University Press, Cambridge, 1970).

\bibitem{note2}  As an illustrative example, let us consider a system of
bronze spheres of diameter $\sigma=0.35~\text{mm}$ ($m=2\times 10^{-7}~\text{
kg}$) with a solid fraction $\phi=3.2\%$, and a pressure $p=3\times 
10^3~\text{Pa}$, recently used in a low gravity experiment [E. Falcon {\em et 
al.}, {\em Phys. Rev. Lett.}:{\bf 83}, 440 (1999)]. This implies 
$\nu_0=0.94~\text{kHz}$.

\bibitem{note}  The model kinetic equation of Ref.\ \onlinecite{BDS99} uses
the exact local homogeneous cooling state of the Boltzmann equation, for
which $f_{\ell }$ is known to be an excellent approximation. In addition, it
uses the exact cooling rate of the Boltzmann equation, which is a bilinear
functional of $f$. If this functional is evaluated using $f_{\ell }$, then
the form for $\zeta $ used here is obtained. Again, this appears to be a
good approximation. Confirmation of the quantitative agreement between the
kinetic model and the Boltzmann equation for the state considered here is
given in Section \ref{sec5}.

\bibitem{USF}  A. Santos and V. Garz\'o, {\em Physica A}:{\bf 213}, 409
(1995); V. Garz\'o and A. Santos, {\em ibid.}:{\bf 213}, 426 (1995).

\bibitem{Bird}  G. A. Bird, {\em Molecular Gas Dynamics and the Direct
Simulation of Gas Flows} (Clarendon, Oxford, 1994).

\bibitem{MGS00}  J. M. Montanero, A. Santos, and V. Garz\'o, {\em Phys.
Fluids}:{\bf 12}, 3060 (2000) and cond-mat/0003364; J. M. Montanero, M.
Alaoui, A. Santos, and V. Garz\'o, {\em Phys. Rev. E}:{\bf 49}, 367 (1994).

\bibitem{BC98}  J. J. Brey and D. Cubero, {\em Phys. Rev. E}:{\bf 57}, 2019
(1998).

\bibitem{Cordero}  R. Ram\'{\i}rez, D. Risso, R. Soto, and P. Cordero, {\em 
Phys. Rev. E}:{\bf 62}, 2521 (2000).
\end{references}
\end{document}